\documentclass[12pt]{article}
\usepackage[usenames, dvipsnames]{xcolor}
\usepackage{jcapmod}

\usepackage{tocloft}
\usepackage[normalem]{ulem}
\usepackage[]{todonotes}
\usepackage{verbatim}
\usepackage{mathrsfs}
\usepackage{cleveref}
\usepackage[margin=.8in]{geometry}
\usepackage[utf8]{inputenc}

\usepackage{multirow}

\usepackage{caption}
\usepackage{longtable}
\usepackage{afterpage}
\usepackage{tikz}
\usepackage{setspace,caption}
\usepackage{lipsum}
\usetikzlibrary{matrix,arrows,calc}
\usepackage{bbm}
\usepackage{slashed}
\usepackage{graphicx}
\usepackage{subcaption}
\usepackage{psfrag}
\usepackage{tensor}
\usepackage{fouridx}
\usepackage{enumitem}
\usepackage{longtable}

\DeclareUnicodeCharacter{200B}{}

\usepackage{bm}
\usepackage{mdframed}
\usepackage{multirow}
\usepackage{soul}
\usepackage{bbold}
\usepackage{multicol}
\usepackage{tikz-cd}
\usepackage{rotating}

\usetikzlibrary{arrows}

  \usepackage{mdframed}
\tikzset{
commutative diagrams/.cd,
arrow style=tikz,
diagrams={>=latex}}

\usepackage{amsmath,amsthm}
\usepackage{amssymb}
\usepackage{mathtools}

\usepackage{feynmf}
\usepackage{marvosym}

\usepackage{import}

\usepackage[utf8]{inputenc}
\usepackage{amsmath}
\usepackage{tikz}
\usepackage{extpfeil}
\usepackage{amsfonts}
\usepackage{amssymb}
\usepackage{bm}
\usepackage{graphicx}
\usepackage{array}
\usepackage{lipsum}
\usepackage{float}
\usepackage{multirow}
\usepackage{hhline}
\usepackage{tabularx}
\usepackage{graphicx}
\usepackage{afterpage}
\usepackage{longtable}
\usepackage{epstopdf}
\usepackage{ytableau}
\usetikzlibrary{decorations.markings}

\usepackage[titletoc]{appendix}
\makeatletter
\newtheorem*{rep@theorem}{\rep@title}
\newcommand{\newreptheorem}[2]{%
\newenvironment{rep#1}[1]{%
 \def\rep@title{#2 \ref{##1}}%
 \begin{rep@theorem}}%
 {\end{rep@theorem}}}
\makeatother
\newtheorem{lemma}{Lemma}[subsection]
\newreptheorem{lemma}{Lemma}

\newtheorem{prop}[lemma]{Proposition}
\newtheorem{conj}[lemma]{Conjecture}
\newreptheorem{conj}{Conjecture}

\theoremstyle{definition}

\theoremstyle{remark}
\newtheorem*{rem*}{Remark}

\newcommand{\jht}[1]{{}}
\newcommand{\jt}[1]{{}}

\definecolor{cobalt}{RGB}{44, 98, 120}
\definecolor{celadon}{rgb}{0.67, 0.88, 0.69}
\definecolor{dm}{cmyk}{.20, 0, .30, 0}
\definecolor{burgundy}{rgb}{0.5, 0.0, 0.13}
\definecolor{plotBlue}{RGB}{94, 130, 181}

\newcommand*\xoverline[2][0.75]{
    \sbox{\myboxA}{$\m@th#2$}
    \setbox\myboxB\null
    \ht\myboxB=\ht\myboxA
    \dp\myboxB=\dp\myboxA
    \wd\myboxB=#1\wd\myboxA
    \sbox\myboxB{$\m@th\overline{\copy\myboxB}$}
    \setlength\mylenA{\the\wd\myboxA}
    \addtolength\mylenA{-\the\wd\myboxB}
    \ifdim\wd\myboxB<\wd\myboxA
       \rlap{\hskip 0.5\mylenA\usebox\myboxB}{\usebox\myboxA}%
    \else
        \hskip -0.5\mylenA\rlap{\usebox\myboxA}{\hskip 0.5\mylenA\usebox\myboxB}%
    \fi}
\makeatother

\begin{document}

\newcommand{\main}{.}
\begin{titlepage}

\setcounter{page}{1} \baselineskip=15.5pt \thispagestyle{empty}

\bigskip\

\vspace{2cm}
\begin{center}
{\Huge 
Electric-Magnetic Duality \\ \vspace{.2cm} in a Class of $G_2$-Compactifications of M-theory}
\end{center}

\vspace{1cm}

\begin{center}
James Halverson$^{1,2}$, Benjamin Sung$^{1,4}$, Jiahua Tian$^{3}$

\vspace{1 cm}
\emph{$^1$Department of Physics, Northeastern University \\ Boston, MA 02115, USA}
\vspace{.3cm}

\emph{$^2$The NSF AI Institute for Artificial Intelligence \\ and Fundamental Interactions}
\vspace{0.5cm}

\emph{$^3$The Abdus Salam International Centre for Theoretical Physics \\ Strada Costiera 11, 34151, Trieste, Italy}
\vspace{.3cm}

\emph{$^4$Department of Mathematics\\University of California, Santa Barbara, CA 93106, USA}
\vspace{.3cm}
\end{center}

\vspace{1cm}
\noindent

\begin{abstract}

We study electric-magnetic duality in compactifications of M-theory on twisted connected sum (TCS) $G_2$ manifolds via duality with F-theory. Specifically, we study the physics of the D3-branes in F-theory compactified on a Calabi-Yau fourfold $Y$, dual to a compactification of M-theory on a TCS $G_2$ manifold $X$. $\mathcal{N}=2$ supersymmetry is restored in an appropriate geometric limit. In that limit, we demonstrate that the dual of D3-branes probing seven-branes corresponds to the shrinking of certain surfaces and curves, yielding light particles that may carry both electric and magnetic charges. We provide evidence that the Minahan-Nemeschansky theories with $E_n$ flavor symmetry may be realized in this way. The $SL(2,\mathbb{Z})$ monodromy of the 3/7-brane system is dual to a Fourier-Mukai transform of the dual IIA/M-theory geometry in this limit, and we extrapolate this monodromy action to the global compactification. Away from the limit, the theory is broken to $\mathcal{N}=1$ supersymmetry by a D-term.

\end{abstract}

\end{titlepage}

\clearpage

\tableofcontents
\newpage

\section{Introduction}
The landscape of M-theory~\cite{Acharya:1998pm, Acharya:2000gb, Witten:2001uq, Acharya:2001gy, Acharya:2004qe} and F-theory~\cite{Vafa:1996xn, Morrison:1996na, 1996morrison, Weigand:2018rez} compactifications gives rise to the largest class of four-dimensional $\mathcal{N} = 1$ string vacua to date. Despite strong coupling effects in general, a supergravity approximation is valid in the large volume limit of a compact background geometry of special holonomy. Additional non-perturbative effects can then be captured by appropriate branes wrapping various cycles, which may or may not be calibrated. Altogether, there is a tight link between the low energy supersymmetric effective physics in four dimensions and the geometry of high-dimensional Ricci flat manifolds. 

Nevertheless, our understanding of M-theory compactifications on $G_2$-manifolds stands in stark contrast to F-theory compactifications on elliptic Calabi-Yau fourfolds. For one, we have no fundamental description of M-theory, while F-theory can certainly be defined as type IIB supergravity with a gauged $SL(2,\mathbb{Z})$ duality group, and with background $D3$ and $(p,q)$ $7$-branes. In such a vein, we have no similar understanding of the duality group in M-theory; the type IIB supergravity action can be written in a manifestly $SL(2,\mathbb{Z})$-covariant fashion, while a similar property is not known for eleven-dimensional supergravity. Finally, we only have a preliminary understanding of weak coupling limits in M-theory on compact $G_2$-manifolds~\cite{Braun:2019wnj}, while there is a systematic understanding in a special class of F-theory compactifications via the Sen limit, which allows us to check various computations with those of perturbative type IIB string theory.  



Our understanding of the respective geometries is plagued with a similar dichotomy. Despite the lack of an explicit bound and a classification, the birational geometry of elliptic Calabi-Yau fourfolds is relatively well-understood via the minimal model program. In particular, working within the category of algebraic varieties gives us a precise control of the K\"{a}hler cone and more refined linear and homological structures beyond just the cohomology of the underlying topological space. On the other hand, the state-of-the-art results for the geometry of compact $G_2$-manifolds pale in comparison, primarily due to the lack of analogous algebraic techniques. There is no analogous classification program, finiteness bounds on the cohomology, or a wealth of constructions with singularities of varying co-dimension. Similarly, there are no clear finiteness and polyhedral conjectures for the $G_2$ analog of the K\"{a}hler cone, part of which can be attributed to the fact that calibrated submanifolds do not necessarily stay calibrated upon deformations in $G_2$-moduli space.  

This state of affairs provides an excellent arena for physical insight to inform the geometry of compact $G_2$-manifolds. Some of the most exciting questions involve physical structures that have historically been surprising in the geometry of Calabi-Yau threefolds, namely, the complexification of $G_2$-moduli space and the realization of singularities leading to non-abelian gauge dynamics and chiral matter. Such a line of inquiry has proven to be fruitful; for example, the authors in~\cite{2016halverson} conjectured an analogue 
of the K\"{a}hler cone for compact $G_2$-manifolds and a general scheme to realize singularities leading to $SU(2)$-gauge enhancements based on general, physical grounds. Conversely, recent substantial progress on constructions of new compact $G_2$-manifolds, such as the twisted connected sum construction~ \cite{2000kovalev, Corti_2013, 2015chnp}, presents a concrete setting to interpret the corresponding physics~\cite{Halverson:2014tya, 2016halverson, 2017guio} and to test new conjectures~\cite{Braun:2018fdp}.
For M-theory compactified on a specific class of twisted connected sum $G_2$-manifolds, \cite{2018} established a detailed chain of dualities relating the effective physics with that of F-theory compactified on a class of elliptic Calabi-Yau fourfolds. 

This duality provides a natural playground to explore the $G_2$-analogues of well-understood F-theory phenomena and leads us to the main subject of our paper: How do the $D3$-brane dynamics and $SL(2,\mathbb{Z})$-monodromies dualize to the geometry of $G_2$-manifolds? Emboldened by this global duality with F-theory, we will conjecture an extrapolation of the M-theory dual results of the local D3-brane dynamics to the global compactification, which we would normally be somewhat more hesitant to do based on the complexities of the gravitational couplings.

\subsection{Summary of results}
Given an M-theory compactification on a $G_2$-manifold $X$, we wish to identify codimension-$2$ singular loci in the complexified moduli space and their associated monodromy action on charged states of the theory. To perform this analysis explicitly, we specialize to the chain of dualities in~\cite{2018} and identify the M-theory duals of $D3$-brane monodromies around $7$-branes. Such M-theory models are realized by a twisted connected sum $(Z_\pm,\varphi)$, with asymptotically cylindrical ends $Z_\pm$ and gluing $\varphi$, with each factor additionally admitting an elliptic fibration. One factor, $Z_-$, is fixed in all models and the associated $K3$-fibration contains $12$ reducible K3-fibers. The singular loci we study corresponds to the contraction of each of these $12$ components to a point, and we study the precise correspondence with the dual D$3$-brane physics.

Our main results and organization are as follows:
\begin{enumerate}
\item
In section~\ref{sec:tcs}, we review the twisted connected sum construction and the chain of dualities established in~\cite{2018}. Specifically, we describe explicitly a class of G2-compactifications with a Higgsable $E_n$-gauge symmetry dual to an F-theory model with base $\mathbb{P}^1 \times \mathbb{P}^1$ with an $E_n$ $7$-brane stack.

\item
In section~\ref{sec:zminus}, we focus on a single building block $Z_-$, which will give one half of the twisted connected sums in all our models. This will be the relevant half for the $D3$-brane dynamics, and we explain in detail the existence of a contraction of a reducible component in the reducible $K3$-fibers in $Z_-$. 
\item
In sections~\ref{sec:general} and~\ref{sec:singlimit}, we review the global and local aspects of the M-theory compactification. We demonstrate that in a local limit, the results of~\cite{2018} reduces to the two dual realizations of the coulomb branch of the $S^1$ reduction of the $5d$ $E_n$ SCFT via the moduli of a $D3$-brane in F-theory and via type IIA compactified on a local $CY3$ with a shrinking generalized del Pezzo surface. We identify the limit of a D3-brane colliding with an $E_n$ 7-brane stack with the limit where one of the 12 reducible components contracts to a point. 
\item
\textbf{Section~\ref{sec:sl2z} serves as our main result}.
We review the correspondence between the lattice of $3-7$ string states on a $D3$-probe with the integral cohomology of a del Pezzo surface. We utilize this correspondence to conjecture the M-theory dual of the monodromy action of a $D3$-probe traversing around an $E_n$ $7$-brane stack and conjecture the generalization to the global compactification.
\item
In section~\ref{sec:breaking}, we remark and demonstrate that the $\mathcal{N} = 2$ to $\mathcal{N}=1$ breaking via the finite Kovalevton is induced by a D-term breaking.
\item
In section~\ref{sec:multipleD3}, we comment on the generalization of our proposal to the case of multiple coincident $D3$-branes.
\end{enumerate}

We believe that our results in section~\ref{sec:sl2z} should hold for any $G_2$-manifold $M \rightarrow S^3$ fibered by $K3$-surfaces and exhibiting a semi-stable degeneration at various points in the base. In the corresponding M-theory compactification, we may consider the physics in the vicinity of a contractible component of a reducible $K3$-fiber, which is isomorphic to a local neighborhood of a del Pezzo surface $dP_n$. Our results in section~\ref{sec:sl2z} carry over in the local physics, and while we used the duality with the $D3$-brane physics in F-theory to justify lifting the monodromy to the global compactification, we believe that this should be more general.

\section{Twisted connected sums and a chain of dualities}\label{sec:tcs}

In this section we will review basic mathematical and physical facts that will be useful in our construction. In Section~\ref{sec:TCS} we will review the twisted connected sum (TCS) construction of $G_2$ manifold. In Section~\ref{sec:duality} we will review the chain of dualities that will be important for our discussion in the following sections and the construction of a special class of TCS $G_2$ manifolds which we will be our main focus. In particular we will review how non-Abelian gauge symmetries arise in this class of geometries. In Section~\ref{sec:localdual} we will review a duality between M-theory on the product of $S^1$ and a local CY3 and D3-brane probing 7-branes which will be important for our later discussions.

\subsection{Twisted connected sum construction and the Kovalev limit}\label{sec:TCS}

In this section we review the basics of twisted connected sum (TCS) construction of $G_2$ manifolds and introduce the notion of Kovalev limit where both the geometry and the physics simplify. 

Denote by $Z_{\pm}$ two threefolds admit the $K3$-fibration structure
\begin{align*}
	K3\hookrightarrow Z_{\pm}\xrightarrow{\pi_{\pm}}\mathbb{P}^1
\end{align*}
with first Chern class $c_1(Z_{\pm}) = [S_\pm]$ where $[S_{\pm}]$ is the class of the generic $K3$ fiber. The threefolds $Z_\pm$ are called \textit{building blocks}. We require $H^3(Z_\pm,\mathbb{Z})$ be torsion-free. Consider the following map
\[
\rho_{\pm} \colon H^2(Z_{\pm},\mathbb{Z}) \rightarrow H^2(S_{0\pm},\mathbb{Z}) \cong \Lambda\equiv U^{\oplus 3}\oplus (-E_8)^{\oplus 2}
\]
induced as the natural restriction maps, where $S_{0\pm}$ is a generic smooth $K3$ fiber over a point $p_{0\pm}\in\mathbb{P}^1$ base of $Z_{\pm}$. We further require $N_{\pm}\coloneqq \text{im}(\rho_{\pm}) \subset H^2(S_{0\pm},\mathbb{Z})$ be primitive in $H^2(S_{0\pm})$, i.e., $\Lambda/N_\pm$ is torsion-free and $T_{\pm}\coloneqq N_{\pm}^\perp \subset \Lambda$.

From $Z_\pm$ we construct two asymptotically cylindrical CY3 (aCyl)
\begin{align*}
	X_{\pm} = Z_{\pm}\backslash S_{0\pm}
\end{align*}
which asymptotes to $S_{\pm}\times [0,T] \times S^1_{b\pm}$. A $G_2$ manifold can then be obtained by gluing $X_{\pm}\times S^1_{e\pm}$ via the TCS construction. In these asymptotic regions $X_{\pm}\times S^1_{e\pm}$ are glued via identifying $S^1_{e\pm}$ with $S^1_{b\mp}$ while the asymptotic $K3$ fibers are mapped to each other by the following hyperk\"ahler rotation called \emph{Donaldson matching}
\begin{align*}
g^* \colon \omega_{S_\pm} &\leftrightarrow Re(\Omega_{S_\mp})\\
g^* \colon Im(\Omega_{S_\pm}) &\leftrightarrow -Im(\Omega_{S_\mp})
\end{align*}
where $\omega$ is the K\"ahler form and $\Omega = Re(\Omega)+iIm(\Omega)$ is the holomorphic two-form of the $K3$ fiber. We will be mainly focusing on the so-called \textit{orthogonal matching} satisfying the following condition 
\[
N_{\pm} \otimes \mathbb{R} = (N_{\pm} \otimes \mathbb{R}) \cap (N_{\mp} \otimes \mathbb{R}) \oplus (N_{\pm} \otimes \mathbb{R} \cap T_{\mp} \otimes \mathbb{R}).
\]
A typical example of orthogonal matching is given by the following lattices
\begin{equation}\label{eq:lattice}
\begin{split}
T_+ &= E_8 \oplus E_8 \oplus U_2 \oplus U_3 \qquad N_+ = U_1\\
T_- &= U_1 \oplus U_2 \qquad\qquad\ \qquad N_- = E_8 \oplus E_8 \oplus U_3
\end{split}
\end{equation}
where we see that in particular $N_+\cap N_- = 0$.

For physical applications, in particular in order to read off the spectrum it is important to know the cohomology of the TCS $G_2$ manifold $M$ and it was given as follows \cite{2015chnp}
\begin{align}\label{eq:cohomology}
\begin{split}
	H^1(M,\mathbb{Z}) =& 0 , \\
	H^2(M,\mathbb{Z}) =& N_+\cap N_- \oplus K_+ \oplus K_- , \\
	H^3(M,\mathbb{Z}) =& \mathbb{Z}[S] \oplus \Gamma^{3,19}/(N_++N_-) \oplus (N_-\cap T_+) \oplus (N_+\cap T_-) \\
	& \oplus H^3(Z_+) \oplus K_+ \oplus H^3(Z_-) \oplus K_- , \\
	H^4(M,\mathbb{Z}) =& H^4(S) \oplus (T_+\cap T_-) \oplus \Gamma^{3,19}/(N_- + T_+) \oplus \Gamma^{3,19}/(N_+ + T_-) \\
	& \oplus H^3(Z_+) \oplus K_+^* \oplus H^3(Z_-) \oplus K_-^* , \\
	H^5(M,\mathbb{Z}) =& \Gamma^{3,19}/(T_+ + T_-) \oplus K_+ \oplus K_+ .
\end{split}
\end{align}
where $K_{\pm} = ker(\rho_{\pm})/[S_{\pm}]$. For a TCS $G_2$ with an orthogonal matching of the type described in Eq.~\ref{eq:lattice} the $U(1)$'s arise from $H^2(M,\mathbb{Z}) = K_+\oplus K_-$ and we will see in Section~\ref{sec:duality} that for the class of TCS $G_2$ manifolds studied in this work we have $|K_-| = 12$ therefore there are always 12 $U(1)$'s arising from the $Z_-$ building block.

From any CY3 $X$ with holomorphic 3-form $\Omega$ and K\"ahler form $\omega$ we can construct $X\times S^1$ with the torsion-free $G_2$ structure
\begin{equation}\label{eq:G2form_on_CY3}
	\Phi = \gamma d\theta\wedge\omega + \text{Re}(\Omega),\ \star\Phi = \frac{1}{2}\omega\wedge\omega - \gamma d\theta\wedge\text{Im}(\Omega)
\end{equation}
where $d\theta$ is the one-form on $S^1$. Following the recipe given by Kovalev \cite{2000affine} one can give a $G_2$-structure $\Phi_M$ to $M$ by writing down the interpolating $G_2$-structures on $X_{\pm}\times S^1_\pm$. For our purpose we will be interested in studying the so-called \textit{Kovalev limit} where $T\rightarrow\infty$ in which limit one expects the two sectors whose spectrum correspond to $K_\pm$ exhibit $\mathcal{N}=2$ SUSY as the associated geometries become $X_{\pm}\times S^1$ \cite{2017guio}. Indeed from Eq.~\ref{eq:cohomology}, one can see that the $\mathcal{N}=1$ vector multiplets from $K_\pm\subset H^2(M,\mathbb{Z})$ and the $\mathcal{N}=1$ chiral multiplets from $K_\pm\subset H^3(M,\mathbb{Z})$ combine into $\mathcal{N}=2$ vector multiplets in 4D in the Kovalev limit. We will see in Section~\ref{sec:breaking} that the partial breaking from $\mathcal{N}=2$ to $\mathcal{N}=1$ by turning on large but finite $T$ is given by a D-term SUSY breaking mechanism at the leading order and the $\mathcal{N}=2$ dynamics is exact when $T\rightarrow\infty$ which matches the expectation that the $G_2$ holonomy of $M$ reduces to the $SU(3)$ holonomy of $X_\pm\times S^1$ in the Kovalev limit.

\subsection{A global M/F-theory duality}\label{sec:duality}

In this work we will focus on a special class of TCS $G_2$ manifolds whose building blocks $Z_\pm$ are both $K3$ and elliptically fibered. The building blocks $Z_\pm$ are constructed from Weierstrass models over $\mathbb{P}^1\times\widehat{\mathbb{P}^1}$ as follows:
\begin{equation}
	y^2 = x^3 + f_{8,4}(z,\widehat{z})w^4 + g_{12,6}(z,\widehat{z})w^6
\end{equation}
where $[y:x:w]$ are the coordinates of $\mathbb{P}^{2,3,1}$ and $f_{8,4}$ and $g_{12,6}$ are polynomials of the indicated degrees in the coordinates $[z_1:z_2]\times[\widehat{z}_1:\widehat{z}_2]$ of $\mathbb{P}^1\times\widehat{\mathbb{P}^1}$. It is then not hard to see that $Z_\pm$ is an elliptic $K3$ fibration over $\widehat{\mathbb{P}^1}$. For the building block $Z_+$ we take $f_{8,4}$ and $g_{12,8}$ to be generic at this stage which can be described as a hypersurface in the toric ambient space with the following weight system
\begin{center}
\begin{tabular}[h]{ccccccc|c}
	$y$ & $x$ & $w$ & $z_1$ & $z_2$ & $\widehat{z}_1$ & $\widehat{z}_2$ & $P$ \\
	\hline
	3 & 2 & 1 & 0 & 0 & 0 & 0 & 6 \\
	6 & 4 & 0 & 1 & 1 & 0 & 0 & 12 \\
	3 & 2 & 0 & 0 & 0 & 1 & 1 & 6
\end{tabular}
\end{center}
where the last column indicates the degrees of the defining polynomial. This weight system will be useful when we construct concrete toric models for the building block in Section~\ref{sec:example}. Note that for generic $Z_+$ there is no non-abelian gauge theory in the 4D effective theory obtained from M-theory on $X$ \cite{2018}, we will discuss how to achieve non-abelian gauge theory in a moment.

For the building block $Z_-$ we use $K3$ surfaces in the family with $N=U\oplus E_8^{\oplus 2}$ as fibers. More concretely for $Z_-$ we specialize the defining Weierstrass model to be
\begin{align*}
	&f_{8,4}(z,\widehat{z}) = z_1^4z_2^4f_{0,4}(z,\widehat{z}), \\
	&g_{12,6}(z,\widehat{z}) = z_1^5z_2^5g_{2,6}(z,\widehat{z}), \\
	&\Delta_{24,12}(z,\widehat{z}) = z_1^{10}z_2^{10}\Delta_{4,12}(z,\widehat{z}).
\end{align*}
We see immediately that $Z_-$ supports $E_8\times E_8$ singularity along two non-intersecting divisors $z_1=0$ and $z_2=0$ and the $E_8$ singularity worsens at 12 double points $z_1 = g_{2,6} = 0$ and $z_2 = g_{2,6} = 0$. For our purpose the following topological numbers will also be useful \cite{2018}:
\begin{align*}
	h^{11}(Z_-)=31,\ h^{21}(Z_-)=20,\ |N_-|=18,\ |K_-| = 12.
\end{align*}

For M-theory compactified a TCS $G_2$ with the building blocks $(Z_+,Z_-)$ constructed in this way (with generic $Z_+$), it was argued in \cite{2018} that there exists a dual F-theory compactification on an elliptic Calabi-Yau fourfold $Y$ with 12 spacetime-filling D3-branes and trivial $G_4$-flux where $Y$ can be described as a complete intersection in an toric ambient space with the following weight system
\begin{center}
\begin{tabular}[h]{cccccccccc|cc}
	$y$ & $x$ & $w$ & $\widehat{y}$ & $\widehat{x}$ & $\widehat{w}$ & $z_1$ & $z_2$ & $\widehat{z}_1$ & $\widehat{z}_2$ & $W$ & $\widehat{W}$ \\
	\hline
	3 & 2 & 1 & 0 & 0 & 0 & 0 & 0 & 0 & 0 & 6 & 0 \\
	0 & 0 & 0 & 3 & 2 & 1 & 0 & 0 & 0 & 0 & 0 & 6 \\
	6 & 4 & 0 & 0 & 0 & 0 & 1 & 1 & 0 & 0 & 12 & 0 \\
	3 & 2 & 0 & 3 & 2 & 0 & 0 & 0 & 1 & 1 & 6 & 6
\end{tabular}
\end{center}
More concretely the defining polynomials in the ambient space are
\begin{align*}
	&\widehat{W} = -\widehat{y}^2 + \widehat{x}^3 + \widehat{f}_4(\widehat{z})x\widehat{w}^4 + \widehat{g}_6(\widehat{z})\widehat{w}^6, \\
	&W = -y^2 + x^3 + f_{8,4}(z,\widehat{z})xw^4 + f_{12,6}(z,\widehat{z})w^6. 
\end{align*}
Note that the defining polynomial $W$ of $Y$ is very similar to the defining polynomial of $Z_+$. In fact $Y$ can be viewed as the fiber product
\begin{align*}
	Y = Z_+\times_{\widehat{\mathbb{P}^1}}dP_9
\end{align*}
where the common $\widehat{\mathbb{P}^1}$ is the one with coordinates $[\widehat{z}_1:\widehat{z}_2]$ and the elliptic fibration structure of $dP_9$ is described by $\widehat{W} = 0$. The main claim of~\cite{2018} can thus be summarized as follows.
\begin{conj}\label{conj:fmduality}
The following physical theories are equivalent.
\begin{itemize}[leftmargin =*]
\item
M-theory on $X$.
\item
F-theory on $Y$ with $G_4 = 0$ and $12$ D3-branes.
\end{itemize}
\end{conj}

On the M-theory side a non-abelian gauge algebra can be achieved by tuning the Weierstrass model of $Z_+$. In particular we consider a resolution of a tuning of $Z_+$ with the following specializations
\begin{equation}\label{eq:ztuning}
f_{8,4}(z, \hat{z}) = z_1^4 f_{4,4}, \quad g_{12,6}(z,\hat{z}) = z_1^5 g_{7,6}, \quad \Delta_{24,12} = z_1^{10}\Delta_{14,12}
\end{equation}
which we denote by $Z_{E_8}$. This realizes a $K3$-fibration where each $K3$-fiber contains an $E_8$ lattice of $(-2)$ curves. The corresponding lattices (compare with $T_+, N_+$ in Eq.~\ref{eq:lattice} in the generic case) are
\[
T_{E_8} = E_8  \oplus U_2 \oplus U_3 \qquad N_{E_8} = E_8 \oplus U_1 
\]
We assume the existence of a hyperk\"{a}hler rotation identifying the $E_8$-lattice in $N_{E_8}$ with an $E_8$-summand of $N_-$ in~\ref{eq:lattice} yielding a smooth $G_2$-manifold $X_{E_8}$. Moreover, we consider the singular, unresolved limit of $Z_{E_8}$, which we denote by $Z_{E_8,sing}$ and we assume that this singular limit is compatible with the matching. In particular, this limit forces the collapse of an $E_8$ lattice of $(-2)$-curves in $Z_-$ \cite{2018}. We denote the corresponding singular $G_2$-manifold by $X_{E_8, sing}$.
\begin{conj}\label{conj:e8fmduality}
M-theory on $X_{E_8,sing}$ is dual to F-theory on the product $Z_{E_8,sing} \times_{\mathbb{P}^1} dP_9$. In particular, the corresponding low energy effective theory exhibits a Higgsable $E_8$ gauge symmetry.
\end{conj}
The generalization to other gauge symmetries is straightforward. One may simply higgs the $E_8$-symmetry on the F-theory side which yields a deformation of the singularity on $Z_{E_8}$, and we assume that this is compatible with the TCS matching.

\subsection{A local duality}\label{sec:localdual}

Besides the duality between M-theory on compact $M$ and F-theory on compact $Y$ with 12 D3-branes, we will also be working with its local version. As having been discussed in Section~\ref{sec:TCS}, at the Kovalev limit (and the large volume limit) the physics becomes $\mathcal{N}=2$ as the two building blocks decouple and we will focus on the 4D $\mathcal{N}=2$ sector obtained by M-theory compactification on $Z_-\times S^1$. In particular we will study the local physics at one of the 12 double points where the $E_8$ singularity in $Z_-$ worsens. 

It is easy to observe from the defining Weierstrass model of $Z_-$ that at such a point where the $E_8$ singularity worsens, one actually has an $E_8-I_1$ type singularity whose resolution leads to a compact shrinkable surface $V\sim dP_n$  \cite{2013tate, Jefferson:2018irk, 2019fiber}. We will postpone the detailed analysis of the geometry of $Z_-$ and $V$ until Section~\ref{sec:zminus} and focus on the physical duality in this section. 

In the vicinity of such an $E_8-I_1$ point the sevenfold $Z_-\times S^1$ can be approximated by $X_V\times S^1$ where $X_V$ is a local CY3 with a compact shrinkable surface $V$. M-theory on $X_V\times S^1$ leads to a 4D $\mathcal{N}=2$ theory which will be denoted by $\mathcal{T}_V$ which is the circle reduction of a 5D $\mathcal{N}=1$ theory obtained by M-theory on $X_V$. For $V\sim dPn$ this 5D $\mathcal{N}=1$ theory is well-known to be the 5D rank-1 $E_n$ theory \cite{1997ims, 1997extremal}. Hence $\mathcal{T}_V$ is a 4D rank-1 theory with KK modes from the circle reduction. 

For our purpose it is important to realize $\mathcal{T}_V$ can also be engineered as the worldvolume theory of a D3-brane probing an affine 7-brane background. In fact it was conjectured in \cite{Hauer_2000, Mohri:2000wu, 2000affine} that M-theory on $X_V\times S^1$ with $V\sim dP_n$ is dual to D3-brane probing $\widehat{E}_n$ 7-branes. It was also argued in \cite{2000hori, 2022uplane} that $\mathcal{T}_V$ can be viewed as D3-brane probing the Coulomb branch of the 5D rank-1 $E_n$ theory on $\mathbb{R}^4\times S^1$ where it is clear that the extra 7-brane that is responsible for the enhancement from $E_n$ to $\widehat{E}_n$ is due to appearance of the KK modes in the circle reduction. 

To summarize, the following (local) duality will be very useful in our subsequent discussions
\begin{align*}
	\text{M-theory on }X_{V\sim dP_n} \times S^1\longleftrightarrow \text{D3-brane probing }\widehat{E}_n\text{ 7-branes}
\end{align*}

\section{Geometry of $Z_-$}\label{sec:zminus}
As pointed out in section~\ref{sec:tcs} and as will be discuss further in section~\ref{sec:d3g2}, much of the $D3$-brane physics will be entirely encoded in the M-theory dual via the building block $Z_-$. In this section, we will discuss in detail the geometry of $Z_-$, exhibit a particular birational model as a hypersurface in a toric variety, and discuss a physically relevant limit in the K\"{a}hler moduli space of $Z_-$. 

In section~\ref{sec:K3fiber} we discuss the structure of the reducible $K3$ fiber and general aspects of the geometry of the building block $Z_-$. In particular, we point out how the geometry of the reducible fibers encodes the structure of an $SU(2)$ gauge enhancement, in agreement with the results of~\cite{2016halverson}. In section~\ref{sec:example}, we discuss a particular birational model of $Z_-$ as a hypersurface in a toric variety. In section~\ref{sec:contraction}, we discuss a particular realization of the contraction of a component of a reducible $K3$-fiber. 

\subsection{Geometry of the reducible $K3$ fibers}\label{sec:K3fiber}
In this section, we specialize to the case relevant for our M-theory compactification, specifically to the $K3$-fibration $Z_- \rightarrow \mathbb{P}^1 \times \widehat{\mathbb{P}^1}$ given in Section~\ref{sec:duality}. More precisely, we will carefully analyze the geometry of building blocks birational to $Z_-$, one of which will be reviewed at length in section~\ref{sec:example}. In particular, we discuss the structure of natural $5$-cycle fibrations in $Z_- \times S^1$, reminiscent of the general ansatz realized in \cite{2016halverson}.


Many such building blocks share the following properties, as first discussed in~\cite{2018}. Recall that at the 12 double points $z_{1,2} = g_{2,6}(z,\hat{z}) = 0$ of $\widehat{\mathbb{P}^1}$, the base of the singular threefold, the singularity worsens due to the $E_8-I_1$ intersection. After a sequence of resolutions, the generic $K3$ fiber degenerates into $V_1 \cup_E V_2$, consisting of two rational elliptic surfaces $V_1$ and $V_2$ intersecting along an elliptic curve \cite{2014klemm,1997friedman,1997aspinwall} over the $E_8-I_1$ point. 

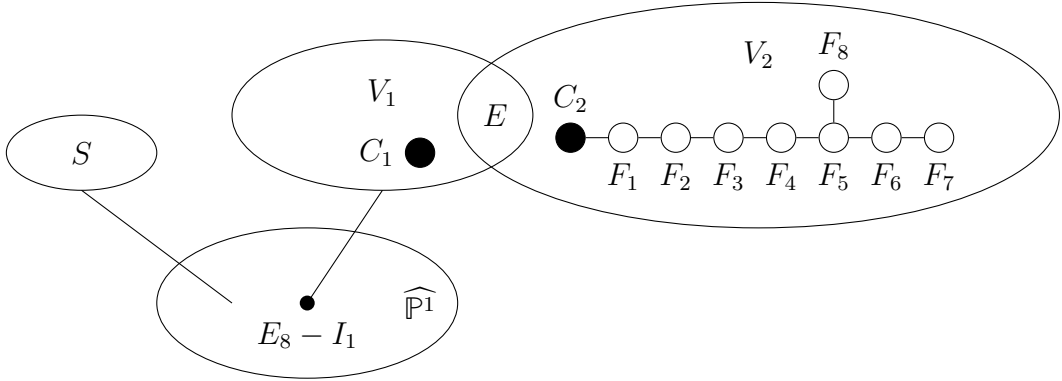
\begin{figure}[h]
\centering
\begin{tikzpicture}
	\draw (0,0) ellipse (2cm and 1cm);
	\node at (1.5,0) {$\widehat{\mathbb{P}^1}$};
	\draw (0,0) node[circle,inner sep=2.0pt,fill,label=below:{$E_8-I_1$}] {};
	\draw (-3,2) ellipse (1cm and 0.5cm);
	\draw (1,2.5) ellipse (2cm and 1cm);
	\draw (6,2.5) ellipse (4cm and 1.5cm);
	\draw (-3,1.5)--(-1,0);
	\node at (-3,2) {$S$};
	\draw (0,0)--(1,1.5);
	\node at (1,2.8) {$V_1$};
	\node at (6,3.3) {$V_2$};
	\node at (2.5,2.5) {$E$};
	\draw (1.5,2) node[fill,minimum size=0.05cm,draw,circle,label=left:{$C_1$}] (b) {};
	\draw (3.5,2.2) node[fill,minimum size=0.05cm,draw,circle,label=above:{$C_2$}] (a) {};
	\draw (4.2,2.2) node[minimum size=0.05cm,draw,circle,label=below:{$F_1$}] (a1) {};
	\draw (4.9,2.2) node[minimum size=0.05cm,draw,circle,label=below:{$F_2$}] (a2) {};
	\draw (5.6,2.2) node[minimum size=0.05cm,draw,circle,label=below:{$F_3$}] (a3) {};
	\draw (6.3,2.2) node[minimum size=0.05cm,draw,circle,label=below:{$F_4$}] (a4) {};
	\draw (7.0,2.2) node[minimum size=0.05cm,draw,circle,label=below:{$F_5$}] (a5) {};
	\draw (7.7,2.2) node[minimum size=0.05cm,draw,circle,label=below:{$F_6$}] (a6) {};
	\draw (8.4,2.2) node[minimum size=0.05cm,draw,circle,label=below:{$F_7$}] (a7) {};
	\draw (7.0,2.9) node[minimum size=0.05cm,draw,circle,label=above:{$F_8$}] (a8) {};
\path[every node/.style={auto=false}]
	(a) [-]	edge (a1)
	(a1) [-] edge (a2)
	(a2) [-] edge (a3)
	(a3) [-] edge (a4)
	(a4) [-] edge (a5)
	(a5) [-] edge (a6)
	(a6) [-] edge (a7)
	(a5) [-] edge (a8);
\end{tikzpicture}\caption{The geometry of $Z_-$ with $k=1$ and $V_2 = gdP_8$. We have $C_1\cdot C_1 = C_2\cdot C_2 = -1$ and $F_i\cdot F_i = -2$. }
\label{fig:geom_Zmi}
\end{figure}

In general, we will work with birational models where the reducible components $V_1, V_2$ are generalized del Pezzo surfaces, denoted $gdP_n$ and $gdP_{18-n}$ respectively. These are similar to del Pezzo surfaces, where $n$ denotes the number of blowups of $\mathbb{P}^2$, but will contain $(-2)$-curves in general. One can flop out a $(-1)$-curve in $gdP_n$ to obtain $gdP_{n-1}$ \cite{1996morrison}. In our case flopping $(-1)$-curves out of $V_2$ $k$ times will lead to the degenerate $K3$ geometry which we denote heuristically by $gdP_{9+k}\cup_E gdP_{9-k}$. The $(-1)$-curve $C_i\subset V_i$ that can be flopped is a rational curve, i.e. a $\mathbb{P}^1$, and therefore, by the degree-genus formula we have $K_{V_i}\cdot_{V_i} C_i = -1$. Thus we have the intersection $E\cdot_{V_i} C_i = 1$ in all flopped phases since $E\in |-K_{V_i}|$ by \cite[Lemma 1.7]{1985kondo}. The geometry of $Z_{-}$ is illustrated in Figure \ref{fig:geom_Zmi}, where for simplicity, we have made a flop of the $(-1)$-curve into the $k=1$ phase and made further flops so that $V_2 = gdP_8$. The $(-2)$-curves are denoted by $F_i$ and intersect along a Dynkin diagram as illustrated, in general. 

As discussed in section~\ref{sec:duality} and in \cite{2018}, the $K$ lattice of $Z_-$ is of rank $12$, i.e. $|K(Z_-)| = 12$. These come from the $12$ reducible $K3$-fibers, each of which consists of $2$ reducible components, arising as in the previous paragraph. Moreover, by a direct computation, one can verify the following intersection relations:
\begin{equation}\label{eq:intersections}
	[V_1] + [V_2] = [S],\quad [C_1]\cdot[V_2] = [C_2]\cdot[V_1] = -[C_i]\cdot[V_i] = 1
\end{equation}
and $[C_i]$ and $[V_i]$ are Poincar\'e dual classes in $H^*(Z_-,\mathbb{Z})$.



From equation~\ref{eq:cohomology}, we deduce that the homology classes of $V_i$ yield non-trivial classes $[V_i \times S^1] \in H_5(X,\mathbb{Z})$ and that the homology classes of $C_i$ yield non-trivial classes $[C_i \times S^1] \in H_2(X,\mathbb{Z})$.  Moreover, their intersection products can be computed away from a general $K3$-fiber, and we conclude that the canonical pairing $H_5(X,\mathbb{Z}) \times H_2(X,\mathbb{Z}) \rightarrow \mathbb{Z}$ induced by Poincar\`e duality with respect to this basis is nothing by the identity matrix. In section~\ref{sec:d3g2}, we will apply these geometric statements to the resulting M-theory compactifications. In analyzing the effective physics, it is convenient and sometimes critical, that the homology classes have calibrated representatives. Thus, we will assume this and note that such an assumption is well supported by existing evidence in the physics literature. 

In~\cite{2016halverson}, a general pattern of $SU(2)$ gauge enhancements in M-theory compactified on a $G_2$-manifold $M$ was conjectured and studied in a number of examples. In general, a $U(1)$ gauge field is obtained in four dimensions via reduction of the $C_3$-field along a $2$-form which is Poincar\`e dual to an integral $5$-cycle $\Sigma_5 \subset M$. A main result of~\cite{2016halverson} was that $\Sigma_5$ should in general admit a fibration by $2$-spheres over a $3$-cycle $[D_{\Sigma_5}] = - [ \Sigma_5 \cap \Sigma_5]$, which was called the Joyce class. Physically, $M2$-branes wrapped on the fibral $2$-spheres correspond to W-bosons on the Coulomb branch of the $SU(2)$ gauge theory which is realized in turn, by collapsing the $2$-spheres to zero volume. In particular, the $U(1)$ gauge coupling is given by the scaling $g^2 \sim \frac{1}{vol(D_\Sigma)}$.


In light of such a general physical ansatz, we will demonstrate how the corresponding fibration structure is realized in our setup. We first recall the geometry of the irreducible component $V_2 \simeq dP_n$ as a fibration over $\mathbb{P}^1$ \cite{1997douglas}. The simplest example of such varieties is $\mathbb{P^1}\times\mathbb{P}^1$ parameterized by homogenous coordinates $([x:y],[s:t])$ where a general anticanonical divisor $-K$ can be written explicitly as
\begin{align*}
	F_{-K}:\ (a_1s^2 + a_2st + a_3t^2)x^2 + (a_4s^2 + a_5st + a_6t^2)xy + (a_7s^2 + a_8st + a_9t^2)y^2 = 0
\end{align*}
In the above form, it is convenient to view the coordinates $[s,t]$ as parametrizing the base $\mathbb{P}^1$, and the coordinates $[x,y]$ as parametrizing the fibral $\mathbb{P}^1$. Fixing a point $[s_0,t_0]$, the fiber intersected with the subvariety $F_{-K}$ yields
\begin{equation}\label{eq:K_covering}
	c_1 s_0^2 + c_2 s_0t_0 + c_3 t_0^2 = 0
\end{equation}
which generically gives two points in the fibral $\mathbb{P}^1$ and one non-reduced point when the above equation degenerates. Thus, we see that the elliptic curve $E\in\mathcal{O}(-K)$ can be viewed as a ramified double covering over the base $\mathbb{P}^1$ branched at the points where Eq.~\ref{eq:K_covering} degenerates. 

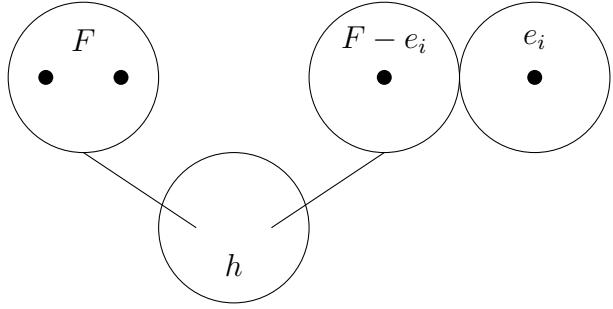
\begin{figure}[h]
\centering
\begin{tikzpicture}
	\draw (0,0) ellipse (1cm and 1cm);
	\node at (0,-0.5) {$h$};
	\draw (-2,2) ellipse (1cm and 1cm);
	\draw (2,2) ellipse (1cm and 1cm);
	\draw (4,2) ellipse (1cm and 1cm);
	\draw (-2,1)--(-0.5,0);
	\node at (-2,2.5) {$F$};
	\draw (0.5,0)--(2,1);
	\node at (2,2.5) {$F-e_i$};
	\node at (4,2.5) {$e_i$};
	\draw (-1.5,2) node[circle,inner sep=2.0pt,fill] {};
	\draw (-2.5,2) node[circle,inner sep=2.0pt,fill] {};
	\draw (2,2) node[circle,inner sep=2.0pt,fill] {};
	\draw (4,2) node[circle,inner sep=2.0pt,fill] {};
\end{tikzpicture}\caption{The geometry of $V_2$. The $\mathbb{P}^1$'s are labeled by their classes in $V_2$. The dots are the intersection points of $E\in\mathcal{O}(-K)$ with the fibral $\mathbb{P}^1$'s in different classes.}
\label{fig:geom_V2}
\end{figure}

A similar picture holds for any del Pezzo surface $dP_n$. Let $h$ denote the pullback of the hyperplane class from $\mathbb{P}^2$, and $e_i$ the exceptional divisors. Fixing an exceptional divisor $e_i$, the linear system $h-e_i$ yields a map $dP_n \rightarrow \mathbb{P}^1$, where we denote the class of the fiber $\mathbb{P}^1$ by $F = h-e_i$. The anti-canonical divisor is then given by
\begin{align*}
	-K = 3h-\sum_{i=1}^{n}e_i.
\end{align*}
Moreover, the fiber class $F$ satisfies $F\cdot F = 0$ with genus $g(F) = 0$, and hence $-K\cdot F = 2$. This can be interpreted as the elliptic curve $E$ in the anticanonical class $-K$ intersecting each fibral $\mathbb{P}^1$ at two points. As a result, $E$ can be viewed as a ramified double covering over the base $\mathbb{P}^1$.
Moreover, we have $-K\cdot e_j = 1$ for each $(-1)$-curve $\mathbb{P}^1$ in the class $e_j$. Therefore at the $n-1$ points of the base $\mathbb{P}^1$ where the fiber $\mathbb{P}^1$ becomes reducible, i.e., becomes $\mathbb{P}^1\cup\mathbb{P}^1$ where one $\mathbb{P}^1$ is in the class $e_j$ and the other $\mathbb{P}^1$ is in the class $F-e_j$, $E$ intersects each $\mathbb{P}^1$ at one point. The geometry of $V_2$ is illustrated in Figure \ref{fig:geom_V2}.

From our discussions above, we have a $\mathbb{P}^1$-fibration $V_i = gdP_n \rightarrow \mathbb{P}^1$. Taking the product $V_i \times S^1 \rightarrow \mathbb{P}^1 \times S^1$, we immediately conclude that the $5$-cycle $V_i \times S^1$ should be identified with the $5$-cycle in the general setting of \cite{2016halverson}. In addition, the base $\mathbb{P}^1 \times S^1$ should be identified with the Joyce class $D_{\Sigma_5}$, and $M2$-branes wrapping the fibral $\mathbb{P}^1$'s should correspond to the W-bosons. On the other hand, we have that the equalities 
\[-[\Sigma_5 \cap  \Sigma_5] = - [V_i \cap V_i \times S^1] = - [K_{V_i} \times S^1],
\]which follows from the intersection relations of~\ref{eq:intersections}. In particular, this is distinctly different from the base $\mathbb{P}^1 \times S^1$ of the fibration of the integral cycle $V_i \times S^1$. The resolution\footnote{We thank Dave Morrison for discussions regarding this point.} is that a general member of the anti-canonical class $-K_{V_i}$ is an elliptic curve which is a double cover of the base $\mathbb{P}^1 \times S^1$, ramified at four points. In particular, the volumes 
\[vol(\mathbb{P}^1 \times S^1) = vol(-K_{V_i} \times S^1)
\] should be identified, and hence correspond to the same gauge couplings.

\subsection{An example}\label{sec:example}
In this section, we discuss an explicit construction of a building block realizing $Z_-$ as a hypersurface in a toric variety. This example was used in the general construction in~\cite{2018}, and we will follow the discussion in~\cite{2017braun}.\footnote{We are grateful to Andreas Braun for sharing initial notes which included discussions on this example.} The ambient toric variety of the singular Weierstrass elliptic fibration with an $E_8 \times E_8$ singularity is given by the following polytope
\begin{align*}
	\Delta_{sing} = \begin{pmatrix}
		-1 & 0 & 2 & 2 & 2 & 0 \\
		0 & -1 & 3 & 3 & 3 & 0 \\
		0 & 0 & -1 & 1 & 0 & 0 \\
		0 & 0 & 0 & 0 & 1 & -1
	\end{pmatrix}
\end{align*}
whose columns correspond to the rays $v_x$, $v_y$, $v_{z_1}$, $v_{z_2}$, $v_{\hat{z}_1}$, $v_{\hat{z}_2}$ where each ray is labeled by its corresponding toric coordinate. The polytope in the $M$ lattice is given by
\begin{align*}
	\nabla = \begin{pmatrix}
		-2 & 1 & 1 & 1 & 1 & 1 \\
		1 & 1 & 1 & 1 & 1 & -1 \\
		0 & 1 & 1 & -1 & -1 & 0 \\
		0 & 0 & -6 & 0 & -6 & 0
	\end{pmatrix}.
\end{align*} 
The hypersurface equation of $Z_{-,sing}$ in the toric ambient space $X_{\Delta_{sing}}$ associated with the polytope $\Delta_{sing}$ is given by
\begin{align*}
	0 = \sum_{m\in\nabla}c_m \hat{z}_2^{\langle m,v_{\hat{z}_2} \rangle}\prod_{v_r\in\Delta_{sing}}r^{\langle m,v_r\rangle+1}.
\end{align*}

The singular variety $Z_{-,sing}$ can be resolved by adding rays to $\Delta_{sing}$ to obtain a new polytope $\Delta$ whose vertices are listed in following matrix
\begin{align*}
	\Delta = \begin{pmatrix}
		-1 & 0 & 2 & 2 & 2 & 0 \\
		0 & -1 & 3 & 3 & 3 & 0 \\
		0 & 0 & -6 & 6 & 0 & 0 \\
		0 & 0 & 0 & 0 & 1 & -1
	\end{pmatrix}
\end{align*}
Among the rays that are needed to desingularize $Z_{-,sing}$, the two rays $v_{z_a} = (0,0,1,0)$ and $v_{z_b} = (0,0,-1,0)$ are more iinteresting than the others for our purpose. For simplicity we could consider the partial resolution of $Z_{-,sing}$ by adding only $v_{z_a}$ and $v_{z_b}$ to $\Delta$. The hypersurface equation of this partially resolved variety in the toric ambient space is
\begin{equation}\label{eq:toric_hypersurface}
	z_az_b\widetilde{P}(x,y,z,\hat{z}) = z_a^2P^b_{12,6}(z,\hat{z}) + z_b^2P^a_{12,6}(z,\hat{z})
\end{equation}
where $P^{a,b}_{12,6}(z,\hat{z})$ are degree 12 in $z$ and degree 6 in $\hat{z}$. Since $v_{z_a}$ and $v_{z_b}$ are never in the same 4D cone of a triangulation of $\Delta$ with additional $v_{z_a}$ and $v_{z_b}$, in the above hypersurface equation we will never have $z_a = 0$ and $z_b = 0$ simultaneously, neither do the pairs $(z_1, z_2)$, $(z_1,z_a)$ or $(z_2, z_b)$ as can be seen from the triangulation of the toric fan. 

Away from the roots of $P^{a,b}_{12,6}(z,\hat{z}) = 0$ on $\widehat{\mathbb{P}^1}$ Eq.~\ref{eq:toric_hypersurface} becomes 
\begin{align*}
	z_az_b\tilde{P} = C_1z_a^2 + C_2z_b^2
\end{align*}
where $C_1$ and $C_2$ are non-zero complex numbers. It is easy to see that over these points the both $\{z_a = 0\}$ and $\{z_b = 0\}$ are empty due to the SR ideal. The $K3$ fiber over generic point of $\widehat{\mathbb{P}^1}$ is thus irreducible. 

The geometry is quite different over points that are solutions to $P^{a,b}_{12,6}(z,\hat{z}) = 0$ on $\widehat{\mathbb{P}^1}$. As the labels $a$ are $b$ are symmetric we focus on the solutions of $P^{a}_{12,6}(z,\hat{z}) = 0$ denoted by $\hat{z}^a_i$. Generically $P^b_{12,6}(z,\hat{z}^a_i) \neq 0$. It is easy to see that $z_b\neq 0$ over $\hat{z}^a_i\in \widehat{\mathbb{P}^1}$ as $z_b = 0$ will necessarily require $z_a = 0$ which violates the SR ideal. Therefore over $\hat{z}^a_i$ we have
\begin{align*}
	0 = z_a\left( \widetilde{P} - z_a P^b_{12,6} \right).
\end{align*}
Thus the $K3$ fiber over $\hat{z}^a_i$ splits to two irreducible components $\{z_a = 0\}$ and $\{\widetilde{P} - z_aP^b_{12,6} = 0\}$. It is easy to see that the component $\{z_a = 0\}$ is toric while the component $\{\widetilde{P} - z_aP^b_{12,6} = 0\}$ is not, hence the volume of $\{z_a = 0\}$ can be controlled by blowing-up or down the toric divisor $v_{z_a}$. This component is nothing but $V_1$ defined in Section~\ref{sec:zminus}. Note that the above arguments hold under the exchange of labels $a$ and $b$ as well.

\subsection{Contraction of a reducible component}\label{sec:contraction}
One of the critical physical limits that we will discuss in section~\ref{sec:d3g2} will be realized geometrically by the contraction of a component of a reducible $K3$-fiber in a birational model of $Z_-$. Thus, it is essential to provide an explicit model of $Z_-$ in which we can realize this limit through either a birational contraction, or through a limit in the K\"{a}hler cone.

Instead of studying an explicit birational contraction of a del Pezzo surface, $dP_n$, in a semistable degeneration of $K3$-surfaces, we will study limits in the K\"{a}hler cone contracting $dP_n$ to a point. Let $\pi \colon \mathcal{Y} \rightarrow B$ be a semi-stable degeneration of $K3$-surfaces with central fiber $\mathcal{Y}_0 = \pi^{-1}(0)$. By \cite{1985kondo}, we may assume that $\mathcal{Y}_0$ is a strict normal crossings of generalized del Pezzo surfaces, $V_1 = gdP_n$ and $V_2 = gdP_{18-n}$ with $n < 9$, intersecting along an elliptic curve contained in the anti-canonical class of both surfaces. We denote by $K(\mathcal{Y})$ the K\"{a}hler cone of $\mathcal{Y}$. The main goal of this section is to discuss the following claim, and we defer the full proof to appendix~\ref{appendix}
\begin{lemma}\label{lem:contraction}
There exists a projective model of $\mathcal{Y}$ and a K\"{a}hler class $J \in K(\mathcal{Y})$ satisfying the following conditions:
\begin{enumerate}[leftmargin=1.25cm]
\item
$J^2 \cdot V_1 = 0$.
\item
$J^3 \neq 0$.
\item
$J^2 \cdot C_1 = 0$ for $C_1 \subset V_1$ a $(-2)$ curve.
\item
If $J^2 \cdot C_2 = 0$ with $C_2 \subset V_2$ a $(-2)$ curve, then $V_1 \cdot C_2 = 0$.
\end{enumerate}
\end{lemma}

We briefly discuss the physics related to the conditions in Lemma~\ref{lem:contraction}. Assume that $\mathcal{Y}$ is a semi-Fano building block for a $G_2$-manifold in the sense of \cite{Corti_2013, 2015chnp}. Assuming that Lemma~\ref{lem:contraction} holds for such a model, we expect that such a limit in K\"{a}hler moduli space also exists for the associated asymptotically cylindrical Calabi-Yau threefold $Z$. In the context of type IIA compactified on $Z$, a K\"{a}hler class satisfying conditions~$(1)$ and $(2)$ contracts a surface component $V_1 = gdP_n$, preserving the overall dimension of $Z$, realizing a $5$d SCFT. As $J$ satisfies condition~$(3)$, the SCFT has at least an $E_n$ flavor symmetry, and by condition $(4)$, the flavor symmetry is precisely given by $E_n$. 


In appendix~\ref{appendix}, we will produce an explicit semi-Fano building block satisfying Lemma~\ref{lem:contraction}. Such a model will be a minor modification of the example in section~\ref{sec:example}, and for now, we will discuss the critical aspects as well as an equivalent formulation of the conditions in Lemma~\ref{lem:contraction}. As discussed, the central fiber $\mathcal{Y}_0$ consists of $V_1, V_2$ intersecting along an elliptic curve. Moreover, we note that there are $n$ and $18 -n$ $(-2)$ curves in $V_1$ and $V_2$ respectively, which are joined by two $(-1)$ curves distributed between the two components and intersecting along a point. In appendix~\ref{appendix}, we will find a toric realization of such a diagram and flop structure. 

Let $X$ be the ambient toric variety and $Y \subset X$ the class of the semi-Fano building block. Let $V'$ be a toric divisor such that $V_1\cap V'|_Y$ and $V_2 \cap V'|_Y$ be the two mentioned $(-1)$-curves. We note that it suffices to check the following four conditions.
\begin{enumerate}[leftmargin=1.5cm]
	\item $J\cdot V_1\cdot V_2\cdot Y = 0$
	\item $J^3\cdot Y \neq 0$
	\item $J\cdot V_1\cdot V'\cdot Y = 0$
	\item $J\cdot V_2\cdot V'\cdot Y \neq 0$
\end{enumerate}
Indeed, the first and second are completely equivalent to the respective first and second conditions in Lemma~\ref{lem:contraction}. The third guarantees that the $(-2)$-curves in $V_1$ cannot be flopped into $V_2$ and hence must be contracted in the limit as $V_1$ contracts to a point. The final condition guarantees that there cannot be additional $(-2)$-curves in $V_2$ that can be flopped into $V_1$ before contracting.

\section{Duals of $D3$-branes in $G_2$-compactifications}\label{sec:d3g2}

This section comprises the main results of the paper. In section~\ref{sec:general} we make some general remarks regarding the D3-brane physics dual to M-theory on $M$ constructed in section~\ref{sec:duality} and global aspects of the $G_2$-comapctification. In section~\ref{sec:singlimit}, we formulate our conjectural singular limit. In particular, we discuss the D3-brane position moduli with respect to the 7-branes to support our conjecture. Moreover, we study the consistency of our proposal with the expected field theory arising in the Kovalev limit. In section~\ref{sec:sl2z}, we discuss the corresponding $SL(2,Z)$-monodromy actions in both the local and global settings. In section~\ref{sec:breaking} we study the breaking of $\mathcal{N}=2$ supersymmetry to $\mathcal{N}=1$ induced by the Kovalevton. In section~\ref{sec:multipleD3} we discuss the physics of multiple D3-branes on top of each other. 

\subsection{General remarks}\label{sec:general}
Consider M-theory on a twisted connected sum $G_2$-manifold $M$ that has an F-theory dual, as in the context of sections~\ref{sec:duality} and \ref{sec:localdual}. In the context of conjecture~\ref{conj:fmduality}, our main goal is to identify the M-theory dual of the D$3$-brane sector and the limit when the D$3$-brane collides with various $7$-brane stacks. In this section, we make several remarks on our compact $G_2$-manifold $M$. 

Recall that $M$ admits a twisted connected sum decomposition into asymptotically cylindrical Calabi-Yau threefolds $Z_{\pm}$, which were defined in Section~\ref{sec:duality}. The geometry of $Z_-$ was studied at length in section~\ref{sec:zminus} and its critical property was that it admits a $K3$-fibration with $12$ reducible $K3$-fibers $S_i$. Moreover, this determines the structure of the $2$ and $5$-cycles in $M$; indeed, we always have 
\[
H^2(M;\mathbb{Z}) = H^5(M;\mathbb{Z}) = 12
\]
for any variation of the building block $Z_+$~\cite{2018}.
We first argue that the $D3$-brane sector is controlled precisely by $Z_-$, justifying our analysis in section~\ref{sec:zminus}. Recall that there are $12$ spacetime filling D$3$-branes in F-theory required by tadpole cancellation. These source $12$ $U(1)$'s which are dual to the $12$ $U(1)$ gauge fields $A_i$ in M-theory arising from Kaluza-Klein reduction of the M-theory $C_3$-field
\begin{equation}\label{eq:c3}
C_3 = A_i \wedge \omega_i + \theta_j\wedge \Phi_j
\end{equation}
In the above, $\omega_i$ are the integral $2$-forms Poincar\`{e} dual to the $5$-cycles $V_i = S_i \times S^1 \subset Z_-\subset Y$ \cite{2018}, and $\theta_i$ and $\Phi_i$ are $4$-dimensional pseudo-scalars and integral $3$-forms on $M$, respectively. We note that the $V_i$'s are precisely the $5$-cycles from the reducible fibers discussed in the previous section. Heuristically, this implies that the D3-brane physics should be dictated by the geometry of $Z_-$, while other aspects of the F-theory physics directly depends on the geometry of $Z_+$ by definition of the F-theory Calabi-Yau fourfold as $Y=Z_+\times_{\widehat{\mathbb{P}^1}} dP_9$. Thus in section~\ref{sec:singlimit}, we will focus on the M-theory physics associated with $Z_-$ using the discussion in section~\ref{sec:zminus} in the Kovalev limit. 

In M-theory on $M$, there are also additional nonperturbative states described by $M2$ and $M5$-branes wrapping $2$ and $5$-cycles, respectively. From the worldvolume action of an $M2$-brane wrapping a curve $C \subset M$
\[
S_{M2} = \int\limits_{C \times \mathbb{R}^1} C_3 = \sum\limits_i \int\limits_C \omega_i \int\limits_{\mathbb{R}^1} A_i 
\]
the resulting state in four dimensions has charge $\int\limits_C \omega_i = [C]\cdot_{Z_-} [V_i] $ under the $i$th $U(1)$ gauge field $A_i$. Similarly, as discussed in~\cite{Long:2021lon}, an $M5$-brane wrapping $V_i \times S^1$ $n$-times has charge $n$ under the gauge field dual to the $i$th $U(1)$ in four dimensions. From~\ref{eq:intersections}, we take a basis of $H^2(M,\mathbb{Z})$ generated by a component $V_i \times S^1$ of each of the $12$ reducible $K3$-fibers, and a basis of $H^5(M,\mathbb{Z})$ generated by the $12$ $(-1)$-flopping curves $C_i \subset V_i$. Summarizing, an $M2$-brane wrapping $C_i$ has electric charge $-\delta_{ij}$ under $A_j$, while an $M5$-brane wrapping $V_i \times S^1$ has magnetic charge $\delta_{ij}$ under $A_j$.

We now review the possible singular limits that can be achieved on $M$. As summarized in Section~\ref{sec:duality}, we can achieve non-abelian gauge symmetry by tuning ADE singularities in every $K3$-fiber on the $K3$-fibration of $M$. Concretely, this can be achieved by engineering $Z_+$ with $K3$-fibers carrying a Picard lattice $N_+ \cap N_-$ of $(-2)$-curves such that under the Donaldson matching, we have the condition
\[
N_+ \cap N_- \neq 0
\]
To see this, note that for a curve $C \in N_+\cap N_-$ we have the equalities
\[
\int_C \omega_\pm = \int_C Re(\Omega_\mp) = 0
\]
where the first equality follows from the Donaldson matching, and the vanishing follows by Poincar\`e duality as $\Omega$ is a $(2,0)$-form and $C$ is dual to a $(1,1)$-form. In particular, from such a gluing, every curve in $N_+ \cap N_-$ must be of zero volume in every $K3$-fiber of $M$ and hence $M$ cannot be resolved via deformations preserving the twisted connected sum condition.

As in Section~\ref{sec:duality}, we can tune an $E_8$-gauge symmetry on $M$ by engineering $Z_+$ with an $E_8$-singularity in every $K3$-fiber, and having the condition $N_+ \cap N_- = E_8$. Also, there is a Dynkin diagram of $(-2)$-curves in each of the $12$ reducible $K3$-fibers, and such a gluing automatically contracts an $E_8$ lattice worth of $(-2)$-curves in each of the $12$ components. Similarly, one can tune an arbitrary $ADE$ gauge symmetry via an analogous method, and in the subsequent section, we will describe a further limit of such models realizing the rank-$1$ superconformal theories with $E_n$ flavor symmetry in the M-theory compactification on $M$.

\subsection{D3-brane moduli and the singular limit}\label{sec:singlimit}



In this section we study the physics of the geometry described in Section~\ref{sec:K3fiber} and our proposed singular limit. As discussed in section~\ref{sec:general}, we may reduce to a local limit of conjecture~\ref{conj:fmduality} and \ref{conj:e8fmduality}. On the M-theory side, we will restrict to a local neighborhood, $X_V$, of $Z_-$ around a single component of the $12$ reducible $K3$-fibers in $Z_-$. On the F-theory side, we will restrict to a local neighborhood of a single $D3$-brane probe in the base $\mathbb{P}^1 \times dP_9$ of the elliptic fibration $Y = Z_+ \times_{\mathbb{P}^1} dP_9$. 

The corresponding physics can be described as follows. In the Kovalev limit, the $G_2$ geometry on one side asymptotes to $Z_-\times S^1$ and thus, one can first consider M-theory compactified on $Z_-$ with a further reduction of that 5D $\mathcal{N}=1$ theory on an $S^1$. The resulting 4D $\mathcal{N}=2$ theory, after restricting to the local neighborhood $X_V$ and decoupling the tower of massive Kaluza-Klein modes, can naturally be viewed as type IIA compactified on $X_{V}$, which is dual to the worldvolume theory of a single spacetime-filling D3-brane in a 7-brane background in type IIB \cite{Masayuki_Noguchi_1999, Yamada_2000}. On the M-theory side, the effective physics is an $\mathcal{N} = 2$ $U(1)$-gauge theory, where the $U(1)$ gauge field is sourced by an integral $2$-form dual to the unique compact surface $V \subset X_V$. Similarly, the worldvolume theory of the $D3$-probe is also an $\mathcal{N} = 2$ $U(1)$ gauge theory in the vicinity of a $7$-brane stack.

To engineer more interesting theories, we will consider singular limits, as discussed in the previous subsection. For concreteness, we will assume that the F-theory geometry $Y$ carries an $E_8$ $7$-brane stack, while in the M-theory dual, there is an $E_8$ Dynkin diagram of $(-2)$-curves in $V = gdP8$ calibrated to zero volume. Our central claim, which holds for arbitrary $E_n$ $7$-brane stacks, is that the limit of the $D3$-brane colliding with the $E_8$ $7$-brane stack is precisely the limit in the M-theory dual when the compact surface component $V = gdP8$ is calibrated to zero volume, which can be done by the results of appendix~\ref{appendix}. Indeed, the limiting $D3$-brane theory is well known to be the rank $1$ $E_8$ Minahan-Nemeschansky theory, which coincides with the M-theory dual limit of contracting $V$ by the results of~\cite{1996seiberg,1997ims,1997extremal,1997ganor,2019fiber}. In the rest of this section, we will explore in more detail aspects of this field theory duality, as well as the natural lift to the duality between the two compact geometries.

Away from the singular limit, M-theory on $X_{gdP_n} \times S^1$ is well-known to correspond to the Coulomb branch of the $S^1$-reduced 5d $\mathcal{N} = 1$ $E_n$ theory~\cite{2000hori, 20215d4d, 2022uplane}. All such theories are of rank $1$, which coincides with the $D3$-brane having a $1$-dimensional modulus normal to the $7$-brane stack. Moreover, the singularities of the Coulomb branch correspond precisely to an $E_n$ stack of $7$-branes and an additional $I_1$ associated with the Kaluza-Klein modes from the $S^1$-reduction in the background of the probe $D3$-brane~\cite{Yamada_2000,Mohri:2000wu}. Mass deformations can be realized as birational transformations of the geometry $X_{gdP_n}$, which correspond in the latter case, to deformations of the background $7$-branes. Finally, as pointed out in section~\ref{sec:zminus}, we note that the generalized del Pezzo surface admits a reducible $\mathbb{P}^1$-fibration $gdP_n \rightarrow \mathbb{P}^1$. $M2$-branes wrapped on the $\mathbb{P}^1$-fibers yield W-bosons and their massless limit yields an $SU(2)$ gauge theory with $N_f = n-1$ flavors with bare gauge coupling $g^2 \sim \frac{1}{vol(\mathbb{P}^1)}$ scaling inversely with the volume of the base. Similarly, we note that decoupling an $I_1$-fiber from the $E_n$ $7$-brane stack yields an identical $SU(2)$ gauge theory phase.

Our discussion elided a subtlety regarding the gauge coupling in the singular limit. The gauge coupling in the F-theory frame is dictated by the complex structure of the elliptic curve fibered over the point in the base. In particular, for $n \geq 6$, the coupling should approach a fixed point of the $SL(2,\mathbb{Z})$ $E_n$ monodromy matrix, and is a constant value. On the other hand, in the M-theory frame, the singular limit of $gdP_n$ contracted to a point clearly sends $vol(\mathbb{P}^1)\rightarrow0$ and hence the bare coupling to infinity for any $n$. For a more precise analysis, we note that the \emph{rates} of vanishing of $\text{vol}(gdP_n)$ and $\text{vol}(C)$ play a critical role~\cite{1997}, where $C \subset gdP_n$ is a curve. In five dimensions, we have \cite{Witten_1996, 1997extremal, 1997douglas, 1997}
\begin{align*}
	\phi_D \sim \phi^2
\end{align*}
where $\phi_D\sim\text{vol}(V_2)$, $\phi\sim\text{vol}(C)$. Hence we have
\begin{align*}
	\tau = \partial_\phi \phi_D \sim \epsilon \rightarrow 0.
\end{align*}
On the other hand, in four dimenisons the scaling is modified by worldsheet instanton corrections to $\phi_D\sim \phi$ and hence $\tau = \partial_\phi \phi_D \sim \text{const.}$ \cite{1997}. For $n \leq 5$, the scaling $g^2 \rightarrow \infty$ is $SL(2,\mathbb{Z})$ equivalent to $0$, which is consistent with the fact that the $E_n$ theories are all infrared free. This is particularly clear in the F-theory frame, where all the relevant $7$-brane stacks are perturbative.

Finally, we compare and verify the matching of BPS states between the two field theories in the local limit. We will be content with matching several lower spin states, and in the subsequent section, we will describe a more general correspondence between $3-7$ string states and the integral cohomology of the del Pezzo surface. One can then match the BPS spectrum of the matters, in particular the electrically charged states. As we have argued in section~\ref{sec:K3fiber}, an M2-brane wrapping a curve $C_2$ will become an electrically charged BPS state under the $U(1)$ dual to $[\Sigma_5]$. On the other hand, the $E_8$ Dynkin diagram inside the surface $V_2\simeq gdP_8$ gives rise to the (massive) flavor symmetry of the low energy theory on the D3-brane probe and the weight of the M2-brane wrapping mode on $C_2$ is $(1,0,0,0,0,0,0,0)$ which is the highest weight of $\mathbf{248}$ of $E_8$. Therefore in the dual F-theory picture on the D3-brane we expect to obtain (massive) spin 0 states $(\mathbf{1},\mathbf{248})$ of $U(1)\times E_8$ from the 3-7 strings which are the dual objects of M2-brane wrapping the curves $C^{\text{spin}\ 0} = C_2 + \sum_ia_iF_i$ that satisfy the condition $C^{\text{spin}\ 0}\cdot_{V_2} C^{\text{spin}\ 0} = -1$ and $g(C^{\text{spin}\ 0}) = 0$ \cite{2019fiber}. Moreover one can consider the curves $C^{\text{spin}\ 1}$ with $C^{\text{spin}\ 1}\cdot_{V_2}C^{\text{spin}\ 1} = 0$ and $g(C^{\text{spin}\ 1}) = 0$ which are the spin 1 BPS states. In particular the $(0)$-curve that corresponds to the highest weight state $(0,0,0,0,0,0,1,0)$ of $\mathbf{3875}$ of $E_8$ is \cite{2019fiber} (see Figure \ref{fig:geom_Zmi})
\begin{align*}
	C^{\text{spin}\ 1} = 2C_2 + 2F_1 + 2F_2 + 2F_3 + 2F_4 + 2F_5 + F_6 + F_8.
\end{align*}

Recall that due to the matching condition the curves in $Z_-$ in the $E_8$ lattice in $V_2$ (see Figure~\ref{fig:geom_Zmi}) are forced to shrink to zero volume in which case the $E_8$ symmetry becomes massless. In the limit $\text{vol}(C_2) = 0$ both the spin 0 states $(\mathbf{1},\mathbf{248})$ and the spin 1 states $(\mathbf{1},\mathbf{3875})$ become massless. If we further let $\text{vol}(V_2)=0$ there will be extra magnetically charged state under the $U(1)$ dual to $[\Sigma_5]$. This suggests that in the limit $\text{vol}(V_2)=0$ what we actually have is an SCFT. Note that the electrically charged massless states $(\mathbf{1},\mathbf{248})$ and $(\mathbf{1},\mathbf{3875})$ non-trivially show up in the BPS spectrum of 4D MN $E_8$ theory \cite{2019distler}. Indeed one can compute the BPS states with higher spin and genus along the same line and match those with the string junctions computed in the dual picture in \cite{2019distler}. Therefore it is tempting to conjecture that the 4D MN $E_8$ theory is realized on the D3-brane world volume in the limit $V_2$ shrinking to a point where all M2-/M5-brane wrapping modes become massless. Hence we conjecture
\begin{conj}\label{conj:transversedist}
	The transverse distance between the D3-brane probe and the $E_n$ 7-branes is proportional to $\text{vol}(V_2)$ in the local CY3 $X$.
\end{conj}

Having formulated the above conjecture in the Kovalev limit, we generalize the arguments to the cases with finite Kovalevton and are led to our main conjecture:
\begin{conj}\label{conj:mainconj}
The following theories are equivalent:
\begin{itemize}[leftmargin=*]
\item
M-theory on $X_{E_n,sing}$ in the limit that a surface $S_i \subset X_{E_n,sing}$ is contracted.
\item
F-theory on $Y_{E_n}$ with $G_4 = 0$ and a single $D3$-brane on the $E_n$ singular locus on the base.
\end{itemize}
\end{conj}
The above conjecture is a natural $\mathcal{N}=1$ generalization of the duality between $\mathcal{N}=2$ theories from IIA on $X_-$ and from the 3/7 system described by the mirror of $X_-$. In particular, the strongly coupled nature of the singular limit is consistent with the fact that we have both $M2$ and $M5$-branes wrapping $2$ and $5$-cycles, and hence electric and magnetic states becoming simultaneously massless.

\subsection{Local and global $SL(2,\mathbb{Z})$-monodromies}\label{sec:sl2z}

In the previous subsection, we conjectured the singular limit of M-theory on the compact $G_2$-manifold $X$ dual to the limit of the $D3$-brane colliding with an $E_n$ $7$-brane stack. The goal of this section is to explore the $SL(2,\mathbb{Z})$-monodromy acting on the BPS states induced by circling this singular limit, which we first discuss in the local case of the $4d$ $\mathcal{N} = 2$ theory supported on a $D3$-brane in the vicinity of a $7$-brane stack and its M-theory dual, and then we extrapolate to the global compactification. We first review the correspondence between these BPS states and the K-theory of a corresponding del Pezzo surface, and note that the action induced by a loop around all $7$-branes is realized in K-theory by a tensor product with the canonical bundle. Finally we conjecture a lift of this action to the ambient $G_2$-manifold. 
Our discussion parallels and builds on the results of~\cite{Hauer_2000}, though we mostly follow the notation and results of~\cite{Harder_2019}. 

Let $\pi \colon Y \rightarrow \Delta$ be a local elliptic fibration over a disc $\Delta$, with fixed base point $p \in \partial \Delta$ and $C = \pi^{-1}(p)$, containing an $E_n$ $7$-brane stack, together with an extra $I_1$. This implies, in particular, that the total monodromy is $M =\begin{pmatrix}
1 & n -9 \\
0 & 1
	\end{pmatrix}$. Let $X \supset dP_n$ be a local Calabi-Yau threefold containing a del Pezzo surface $dP_n$ which is contractible to a point. As discussed in the previous subsections, there is an identification between the $4$d $\mathcal{N} = 2$ theory on a D$3$-brane probing the $E_n$ $7$-brane stack in $B$ with the $4$d $\mathcal{N} = 2$ theory obtained from type IIA compactified on $X$. The lattice of BPS states on the D3-brane is described by the relative homology group $H_2(Y,C; \mathbb{Z})$ together with a pairing, which we take to be an integral modification of the pairing defined in \cite{Grassi:2013kha,Grassi:2014ffa,Grassi:2014sda,Grassi:2018wfy}. Fixing a basis $\{v_1, \ldots, v_{n+3} \}$ of $(p,q)$ $7$-branes for the $E_n$ $7$-brane stack, we define the pairing on the basis of $3-7$ strings as follows:
\begin{equation*}
\langle \begin{pmatrix} p_i \\ q_i \end{pmatrix} , \begin{pmatrix} p_j \\ q_j \end{pmatrix} \rangle = \begin{cases}
q_i p_j - p_i q_j&\text{ if } i< j \\
1&\text{ if } i = j \\
0&\text{ if } i > j
\end{cases}
\end{equation*}
which extends to the full lattice by linearity. Finally, we recall that there is an asymptotic charge map $a(J) \colon H_2(Y,C;\mathbb{Z}) \rightarrow H_1(C; \mathbb{Z})$ taking a junction $J$ to the sum of its $7$-brane charges.

The correspondence at the level of the BPS states can be summarized by the following diagram:
\begin{equation}\label{knum}
\begin{tikzcd}[  ar symbol/.style = {draw=none,"\textstyle#1" description,sloped}
,isomorphic/.style = {ar symbol={\simeq}}]
H_2(Y,C;\mathbb{Z}) \arrow[d,isomorphic]\arrow[r,"a(J)"]& H_1(C;\mathbb{Z}) \arrow[d,isomorphic] \\
K_{num}(dP_n) \arrow[r,"i^*"] & K_{num}(E) 
\end{tikzcd}
\end{equation}
where $i \colon E \xhookrightarrow{} dP_n$ is the is the inclusion of an elliptic curve, contained in the anti-canonical class, into $dP_n$. Roughly speaking, the Grothendieck group $K_0(X)$ for $X$, a smooth projective variety, is the class $[F]$ of all coherent sheaves $F$ on $X$ modulo the relation $[F] = [E] + [G]$, if there is an exact sequence $E \rightarrow F \rightarrow G$ of coherent sheaves on $X$. The numerical Grothendieck group $K_{num}(X)$ is then defined as the Grothendieck group modulo the kernel of the Euler pairing $\chi(-,-)$ which is defined as $\chi(E,F) = \sum\limits_i (-1)^i dim(Ext^i(E,F))$.

The critical property for our purposes is that $K_{num}(X)$ is a finite rank lattice, with a canonical pairing given by the Euler pairing. In our case, one may think of $K_{num}(dP_n)$ and $K_{num}(E)$ as simply, the graded integral cohomology rings $H^*(dP_n;\mathbb{Z})$ and $H^*(E;\mathbb{Z})$ with the usual pairing of cycles. In particular, $K_{num}(dP_n)$ is a lattice of rank $n + 3$, in agreement with the total number of $(p,q)$ $7$-branes.

We now discuss the induced $SL(2,\mathbb{Z})$-monodromy on both sides. In the F-theory frame, there is a natural duality induced on the $D3$-probe by traversing a loop around all $(p,q)$ $7$-branes in the base $\Delta$. Such a loop induces a natural action on the relative homology group $H_2(Y,C;\mathbb{Z})$ via Hanany-Witten moves. As an example, assume that $H_2(Y,C;\mathbb{Z}) = \langle v_1,v_2 \rangle$ with $(p,q)$-charges $(p_1,q_1)$ and $(p_2,q_2)$ respectively. Then such a loop induces the actions
\begin{align*}
v_2 \mapsto v_2 - \langle v_1, v_2 \rangle v_1 \qquad v_1 &\mapsto v_1 + \langle v_1,v_2\rangle v_2 \\
&\mapsto v_1 + (\langle v_1,v_2 \rangle(v_2 - \langle v_1, v_2 \rangle v_1)
\end{align*}
We remark that critically, the junction pairing is preserved if and only if the $D3$-brane traverses a loop around \emph{all} the $7$-branes. In particular, the monodromy induces a duality of the theory only under such a loop. Indeed, this is consistent with the general strategy employed in \cite{Grassi:2016bhs,Grassi:2021ptc} where the precise flavor symmetry and matter spectrum on the $D3$-probe was identified by truncating the naive spectrum by a self-duality under a loop around \emph{all} the $7$-branes in a local neighborhood.

By analyzing diagram~\ref{knum}, we should obtain an analogous automorphism of the lattice $K_{num}(dP_n)$ preserving the Euler pairing. We claim that such an action is simply induced by Serre duality via the tensor product with the canonical bundle:
\begin{align*}
K_{num}(dP_n) &\rightarrow K_{num}(dP_n)\\
[E] &\mapsto [E \otimes \omega_{dP_n}]
\end{align*}
We will check that such an action induces precisely the total monodromy $M =\begin{pmatrix}
1 & n - 9 \\
0 & 1
	\end{pmatrix}$ after pullback to $K_{num}(E)$ via diagram~\ref{knum}. Indeed, $K_{num}(E)$ is generated by the classes $[\mathcal{O}_E], [\mathcal{O}_p]$, i.e. the structure sheaf and a skyscraper sheaf, respectively. In terms of the total integral cohomology ring $H^*(E;\mathbb{Z})$, these correspond to the fundamental class and the class of a point, generating $H^0(E;\mathbb{Z})$ and $H^2(E;\mathbb{Z})$, respectively.

We will compute the restriction of the classes $[\mathcal{O}_{dP_n}], [\omega^{-1}_{dP_n}]$ to $K_{num}(E)$ before and after the action, and demonstrate that it coincides with the above monodromy. For simplicity, we work with the corresponding classes in cohomology, which correspond to $(1,0,0), (1, -K_{dP_n}, \frac{1}{2}K_{dP_n}^2) \in H^*(dP_n;\mathbb{Z})$ respectively. Restricting to $E$, these yield the classes $(1,0), (1, 9-n)$ respectively, where for example, the divisor $-K_{dP_n}$ restricts to $9-n$ points on $E$. On the other hand, the tensor product with $\omega_{dP_n}$ yields the classes $[\omega_{dP_n}],[\mathcal{O}_{dP_n}] \in K_{num}(dP_n)$ corresponding to the cohomology classes $(1, K_{dP_n}, \frac{1}{2}K_{dP_n}^2), (1,0,0) \in H^*(dP_n;\mathbb{Z})$ respectively. Restricting to $E$, these yield the classes $(1, n-9), (1,0)$ respectively. Thus, the corresponding action is given by the matrix $\begin{pmatrix} 1 & 0 \\ n-9 & 1 \end{pmatrix}$, which is nothing but our claimed matrix after a change of basis.

From the above two paragraphs, we have found that the monodromy induced by a $D3$-probe traversing a loop around all $7$-branes is dual to the monodromy action
\begin{align*}
H^*(dP_n;\mathbb{Z}) &\rightarrow H^*(dP_n;\mathbb{Z})\\
(d_4,d_2,d_0) &\mapsto (d_4, d_4 K_{dP_n} + d_2, d_0 + d_2 K_{dP_n} + \frac{1}{2}d_4 K_{dP_n}^2)
\end{align*}
Our notation reflects the fact that these classes correspond precisely to the $D0$, $D2$, and $D4$-brane charges in type IIA on the local Calabi-Yau threefold $X$. 

It is now straightforward to conjecture a generalization of this formula to M-theory on a twisted connected sum $G_2$-manifold $M$. Assume that $X$ is an asymptotically cylindrical Calabi-Yau threefold with a contractible del Pezzo surface $dP_n$. In the context of the duality of M-theory on $CY3 \times S^1$ with type IIA on $CY3$, an $M2$-brane wrapped on a $2$-cycle in $CY3$ corresponds to a $D2$-brane wrapped state, and an $M5$-brane wrapped on a $5$-cycle $D \times S^1$ corresponds to a $D4$-brane wrapped on $D$. Thus, we conjecture that the corresponding monodromy acting on a linear combination of states from an $M2$-brane wrapping a curve $C \subset dP_n$ and an $M5$-brane wrapping $dP_n \times S^1$ is simply given by the following
\begin{equation}\label{eq:monodromy}
\begin{split}
H^2(M;\mathbb{Z}) \oplus H^5(M;\mathbb{Z}) &\rightarrow H^2(M;\mathbb{Z}) \oplus H^5(M;\mathbb{Z})\\
(m_5, m_2) &\mapsto (m_5, m_5 K_{dP_n} + m_2)
\end{split}
\end{equation}
where by $m_5 K_{dP_n}$, we mean an integer multiple of the $2$-cycle $K_{dP_n}$, which is the inclusion of the class of the canonical divisor $K_{dP_n} \subset dP_n$ into $M$.

As discussed in section~\ref{sec:general}, the compact $G_2$-manifold $M$ satisfies $H^2(X,\mathbb{Z}) = H^5(X,\mathbb{Z}) = 12$ with a rather simple basis for the intersection pairing. Given an element $(m_5, m_2) \in H^2(X,\mathbb{Z}) \oplus H^5(X,\mathbb{Z})$ with an expansion
\[
m_5 = \sum\limits_i a_i [V_i \times S^1],
\]
where $V_i = dP_n$ with $n \leq 9$, is a fixed component of the $i$th reducible $K3$ fiber, the electric and magnetic charges, $e_i, d_i$ under the $i$th $U(1)$ dual to $V_i \times S^1$ is given by $m_2 \cdot V_i$ and $a_i$ respectively. From equation~\ref{eq:monodromy}, the monodromy associated with circling the singular limit associated with contracting the $i$th reducible component thus acts as
\begin{equation}\label{eq:monodromycharge}
\begin{split}
e_i = m_2 \cdot V_i &\mapsto e_i' = m_2 \cdot V_i + (9-n)a_i \\
d_i = a_i &\mapsto d_i' = a_i
\end{split}
\end{equation}
where we have used that $K_{dP_n} \cdot V_i = 9-n$.

\subsection{Breaking $\mathcal{N}=2$ to $\mathcal{N}=1$}\label{sec:breaking}

Though in the Kovalev limit the low energy effective theory can be well approximated by $\mathcal{N}=2$ theory obtained from compactification of M-theory on $X_-\times S^1$, the 4D theory is actually $\mathcal{N}=1$ for any finite Kovalevton. It is useful to investigate the SUSY breaking mechanism in this process.

In this section for simplicity we consider the case where the 4D theory is described by a Lagrangian. We consider a smooth TCS $G_2$ manifold $M$ with building blocks $(Z_+,Z_-)$ whose $G_2$-structure $\Phi$ can be expanded as
\begin{align*}
	[\Phi] = \sum S^i[\rho_i^{(3)}]
\end{align*}
where $[\rho_i^{(3)}]\in H^3(M,\mathbb{Z})$. Upon compactification the three-form field $C_3$ can be expanded as
\begin{align*}
	C_3 = \sum_I A^I\wedge \omega_I^{(2)} + \sum_i P^i \rho_i^{(3)}
\end{align*}
where $\omega_I^{(2)}\in H^2(M,\mathbb{Z})$. In this notation the scalar component of the 4D chiral multiplet is $\phi^i = -P^i + iS^i$ \cite{2017guio}.

The non-gravitational part of the 4D Lagrangian is
\begin{equation}\label{eq:general_form}
	\mathcal{L}_{NG} = \frac{1}{2}\kappa_{IJk} \left( S^kF^I\wedge\star_4F^J -  P^kF^I\wedge F^J \right) - \frac{1}{2\lambda_0}\lambda_{ij} \left( dS^i\wedge\star_4dS^j + dP^i\wedge\star_4dP^j \right)
\end{equation}
where 
\begin{align*}
	\kappa_{IJk} &= \int_X \omega_I^{(2)}\wedge\omega_J^{(2)}\wedge\rho_k^{(3)}, \\
	\lambda_{ij} &= \int_X \rho_i^{(3)}\wedge\star_{g(\Phi)}\rho_j^{(3)}, \\
	\lambda_0 &= \int_X \Phi\wedge\star_{g(\Phi)}\Phi.
\end{align*}
The bosonic part of the chiral multiplet sector of $\mathcal{L}_{NG}$ is 
\begin{align*}
	\mathcal{L}_{NG,s} &= -\frac{1}{2\lambda_0}\lambda_{ij}\left( \partial_\mu S^i\partial^\mu S^j + \partial_\mu P^i\partial^\mu P^j \right) \\
	&= -\frac{1}{2\lambda_0}\lambda_{ij} \partial_\mu \phi^i\partial^\mu \overline{\phi}^j.
\end{align*}
and the bosonic part of the gauge sector of $\mathcal{L}_{NG}$ is
\begin{align*}
	\mathcal{L}_{NG,g} &= \frac{1}{2}\kappa_{IJk} \left( S^kF^I_{\mu\nu}F^{J,\mu\nu} - P^kF^I\wedge F^J \right) \\ 
	&= \frac{1}{2}\kappa_{IJk} \left( \text{Im}(\phi^k) F^I_{\mu\nu}F^{J,\mu\nu} + \text{Re}(\phi^k) F^I\wedge F^J \right).
\end{align*}
Therefore, up to an overall $1/2$ factor, $\mathcal{L}_{NG}$ can be written as
\begin{equation}\label{eq:LNG}
	\mathcal{L}_{NG} = -\left( \frac{\lambda_{ij}}{\lambda_0}\partial_\mu \phi^i\partial^\mu \overline{\phi}^j - \kappa_{ijk}\text{Im}(\phi^k) F^i_{\mu\nu}F^{j,\mu\nu} \right) + \kappa_{ijk}\text{Re}(\phi^k)F^i\wedge F^j
\end{equation}

Recall that for any $\mathcal{N}=2$ gauge theory the bosonic part of the Lagrangian can be written in the following form \cite{2010gaiotto}:
\begin{align*}
	\mathcal{L} = -\text{Im}\tau_{ij}\left( \partial_\mu \phi^i\partial^\mu \overline{\phi}^i + F^i_{\mu\nu}F^{j,\mu\nu} \right) + \text{Re}\tau_{ij}F^I\wedge F^J.
\end{align*}
Therefore for $\mathcal{L}_{NG,g}$ to be $\mathcal{N}=2$ supersymmetric we must have:
\begin{align*}
	\tau_{ij} = \phi^k\kappa_{ijk}
\end{align*}
and
\begin{align*}
	\text{Im}(\tau_{ij}) = S^k\kappa_{ijk} = -\frac{\lambda_{ij}}{\lambda_0}.
\end{align*}
Hence we require
\begin{equation}\label{eq:N=2SUSY}
	S^k\kappa_{ijk} = -\frac{\lambda_{ij}}{\lambda_0}.
\end{equation}
Therefore for the $K_{\pm}$ sector in the Kovalev limit we expect the following relation to hold on $X_{\pm}\times S^1_{\pm}$ where $X_{\pm} = Z_{\pm}\backslash S_{\pm}$:
\begin{equation}\label{eq:2SUSY}
	S^k\int_{X_{\pm}\times S^1_{\pm}} \omega_i^{(2)}\wedge\omega_j^{(2)}\wedge\rho_k^{(3)} = -\frac{\int_{X_{\pm}\times S^1_{\pm}} \rho_i^{(3)}\wedge\star_{g(\Phi)}\rho_j^{(3)}}{\sum_{a,b}S^aS^b\int_{X_{\pm}\times S^1_{\pm}} \rho_a^{(3)}\wedge\star_{g(\Phi)}\rho_b^{(3)}}
\end{equation}
since in the Kovalev limit one gets $\mathcal{N}=2$ SUSY on $X_{\pm}\times S^1_{\pm}$ \cite{2017guio}. In the Kovalev limit we can also set the dimensionless volume $\lambda_0=1$, i.e., set $\text{vol}(X)$ at its reference volume given by the moduli $S^i\in K_{\pm}$ at their VEVs (Eq.~(2.25) in \cite{2017guio}) since the Kolevton moduli $T$ decouples. With these simplifications to show that the system is $\mathcal{N}=2$ we need to show that $S^k\kappa_{ijk}=-\lambda_{ij}$.

Hence we will focus on the term
\begin{align*}
	\rho_{ab} = \int_{X_{\pm}\times S^1_{\pm}}\rho_a^{(3)}\wedge\star_{g(\Phi)}\rho_b^{(3)}
\end{align*}
where $\rho_i^{(3)} = \omega_i^{(2)}\wedge dt$ where $t$ parameterizes $S^1_{\pm}$. In the Kovalev limit the metric of $X_{\pm}\times S^1_{\pm}$ factorizes as
\begin{align*}
	g(\Phi) = g(X_{\pm})\times g(S^1_{\pm}). 
\end{align*}
Now we have 
\begin{align*}
	\rho_{ab} &= \int_{X_{\pm}\times S^1} \omega_a^{(2)}\wedge dt \wedge \star_{g(X_{\pm})\times g(S^1_{\pm})}\left( \omega_b^{(2)}\wedge dt \right) \\
	&= \int_{X_{\pm}} \omega_a^{(2)}\wedge \star_{g(X_{\pm})} \omega_b^{(2)}
\end{align*}
With a suitable coordinate transformation of the vector space $H^*(X_{\pm},\mathbb{Z})$ we have the following expansion \cite{Strominger:1985ks}:
\begin{align}\label{eq:strominger}
	\star_{g(X_{\pm})} \omega_i^{(2)} = -\omega_i^{(2)}\wedge J + \frac{3\int_{X_{\pm}} \omega^{(2)}_i\wedge J\wedge J}{2\int_{X_{\pm}} J\wedge J\wedge J}J\wedge J.
\end{align}
where $J$ is the K\"ahler form of $X_{\pm}$ and in the Kovalev limit we have (cf. Eq.~\ref{eq:G2form_on_CY3})
\begin{align*}
	\Phi = -J\wedge dt + \text{Re}(\Omega) = \sum S^i\rho^{(3)}_i
\end{align*}
where $\Omega$ is the holomorphic 3-form of $X_{\pm}$ and $dt$ is the 1-form of $S^1$. Thus the non-vanishing part of $\omega^{(2)}_i\wedge J$ can be written as 
\begin{align*}
	\omega^{(2)}_i\wedge J = -S^k\omega^{(2)}_i\wedge\omega^{(2)}_k
\end{align*}
where $\omega^{(2)}_k\in K_{\pm}$ and the lift of $\omega^{(2)}_k\wedge dt$ is in $H^3(X)$. Therefore in the limit $\text{vol}(X_{\pm})\rightarrow\infty$ and thus the second term of equation~\ref{eq:strominger} vanishes, we have
\begin{align*}
	\int_{X_{\pm}\times S^1} \rho_a^{(3)}\wedge\star_{g(\Phi)}\rho_b^{(3)} = -S^k\kappa_{ijk}.
\end{align*}
Hence the RHS of Eq.~\ref{eq:N=2SUSY} becomes (with $\lambda_0$ set to 1)
\begin{align*}
	-\lambda_{ij} = -\rho_{ij} = S^k\kappa_{ijk}
\end{align*}
which is equal to the LHS. Thus we see that $\mathcal{N}=2$ SUSY holds on $X_{\pm}\times S^1_{\pm}$ at the Kovalev limit.

To show that $\mathcal{L}_{NG}$ is broken from $\mathcal{N} = 2$ to $\mathcal{N}=1$ it is sufficient to show that for finite Kovalevton
\begin{equation}\label{eq:breaking_condition}
	S^k\kappa_{ijk} \neq -\frac{\lambda_{ij}}{\lambda_0}
\end{equation}
for any finite $T$.

First we focus on the LHS of Eq.~\ref{eq:breaking_condition}. In general we have
\begin{align*}
	S^k\kappa_{ijk} = \int_{X_{\pm}\times S^1_{\pm}} \omega_i^{(2)}\wedge\omega_j^{(2)}\wedge[\Phi]
\end{align*}
and we will focus on the term
\begin{align*}
	\kappa_{ij[S]} = \int_{X_{\pm}\times S^1_{\pm}} \omega_i^{(2)}\wedge\omega_j^{(2)}\wedge[S].
\end{align*}
where $[S]$ is the Poincar\`e dual to the class of the $K3$-fiber.

We can write the integral $\kappa_{ij[S]}$ in terms of an intersection in $X$
\begin{align*}
	\kappa_{ij[S]} = W_i\cdot_X W_j\cdot_X W_{[S]}
\end{align*}
where $W_{[S]}$ is the homology class corresponds to the $K3$ fiber $S$ and $W_i$ is the 5-cycles dual to $\omega_i^{(i)}$. After restricted to the $K_{\pm}$ sector we can write the above intersection as
\begin{align*}
	\kappa_{ij[S]} = \widetilde{W}_i\cdot_{X_{\pm}\times S^1_{\pm}} \widetilde{W}_j\cdot_{X_{\pm}\times S^1_{\pm}} \widetilde{W}_{[S]}
\end{align*}
where $\widetilde{W}_i$ is the 4-cycle that is the image of $W_i$ under $\widetilde\pi: X\rightarrow Z_{\pm}$. Recall that $Z_{\pm}$ is a $K3$ fibration of $S$ therefore 
\begin{equation}\label{eq:kappa_ijS}
	\kappa_{ij[S]} = \widehat{W}_i\cdot_{S} \widehat{W}_j
\end{equation}
where $\widehat{W}_i$ is the pullback of $\widetilde{W}_i$ under the inclusion $\pi \colon S \xhookrightarrow{} \widehat Z_{\pm}$.

Recall that we have the map
\begin{align*}
	\rho_{\pm}: H^2(Z_{\pm},\mathbb{Z})\rightarrow H^2(S_{\pm},\mathbb{Z})
\end{align*}
which, with Poincar\'e duality, becomes
\begin{align*}
	\rho^{d}_{\pm}: H_4(Z_{\pm},\mathbb{Z})\rightarrow H_2(S_{\pm},\mathbb{Z}).
\end{align*}
As $\omega_i^{(2)}\in \text{ker}\rho_{\pm}$, its Poincare dual $\widetilde{W}_i\in \text{ker}\rho^{d}_{\pm}$ and $\rho^{d}_{\pm}$ is nothing but the pushforward of $\widehat{\pi}$. Therefore in Eq.~\ref{eq:kappa_ijS}, $\widehat{W}_i$'s are trivial 2-cycles on $S$ hence $\kappa_{ij[S]} = 0$.

The above calculation shows that $S^k\kappa_{ijk}$, i.e., $\text{Im}(\phi^k\kappa_{ijk})$ receives no contribution from the moduli $[S]$, hence the Kovalevton. Hence we would expect the gauge sector in Eq.~\ref{eq:LNG}:
\begin{align*}
	\mathcal{L}_{A} = -\text{Im}(\tau_{ij})F^i_{\mu\nu}F^{j,\mu\nu}
\end{align*}
depends only on the data of the compact sector $k_{\pm}\subset X_{\pm}$ and does not depend on the Kovalevton $T$.

It then remains to show that the RHS of Eq.~\ref{eq:breaking_condition} depends on $T$ for any finite value of $T$ which is actually obvious since $\lambda_0$ depends on all the moduli of $Y$, in particular $T$, as given by the following equation (Eq.~(2.25) and (3.25) in \cite{2017guio}):
\begin{align*}
	\lambda_0 \sim V_{K3}(2T+\mathcal{F}(S)) + O(e^{-T})
\end{align*}
where $\mathcal{F}(S)$ is a function of the moduli $S$ other than Kovalevton and an overall volume volume modulus $R$ and $V_{K3}$ is the volume of the K3 fiber. We see that $\lambda_{ij}/\lambda_0$ is inevitably a function of $T$ when the correction is not suppressed for finite $T$. This $T$ dependence breaks the equality of Eq.~\ref{eq:breaking_condition} away from the Kovalev limit as now $\lambda_0 = \lambda_0(T)$ hence breaks the $\mathcal{N} = 2$ SUSY of the $X_{\pm}$ sector at the Kovalev limit as well. In particular we see that in the gauge theory sector it is a D-term breaking mechanism at the leading order as it changes the kinetic coupling of the original $\mathcal{N}=2$ theory.

Certainly the above Lagrangian approach does not apply to strongly coupled physics where a Lagrangian description is missing but one can still assume the theory is partially broken by deforming the original $\mathcal{N}=2$ theory by certain operator in a similar manner described in \cite{Heckman_2011}. Note that all the discussions in this section are based on the $G_2$ geometry hence is on the M-theory side of the duality chain. It is interesting to study the dual of this partial breaking mechanism in the 3/7 system and we will leave this to future study. Here we conjecture the $\mathcal{N}=2$ SUSY in the 3/7 system might be broken by the coupling to gravity for finite Kovalevton.

\subsection{Multiple D3-branes}\label{sec:multipleD3}

In this section we discuss the physics of multiple D3-branes on top of each other near the 7-branes. On the dual F-theory side the physics is quite clear. When $n$ D3-brane are on top of each other the world volume gauge theory is enhanced to $SU(n)$. The W-bosons that are necessary for such enhancement are the 3-3 strings stretching between the D3-branes that become massless in the limit when they coincide.

On the M-theory side the picture is more interesting. Recall that generically there are 12 distinct double points on $\widehat{\mathbb{P}^1}\subset Z_-$ where the $K3$ fiber $S_i$ becomes reducible, i.e., $S_i = V^i_1\cup_E V^i_2$, $i=1,\cdots,12$. As the generic geometry is conjectured to dual to a single D3-brane probing 7-branes, we would naturally look at the geometry when some of the 12 double points coincide on $\widehat{\mathbb{P}^1}$. We assume there are $n$ coincident double points at $p\in\widehat{\mathbb{P}^1}\subset Z_-$. Above $p$ the $K3$ fiber $S_p$ becomes reducible and is again Kulikov type II. We have
\begin{equation}\label{eq:multipleD3_geom}
	S_p = V_0\cup_{C_{0,1}}V_1\cup_{C_{1,2}}V_2\cup_{C_{2,3}}\cdots\cup_{C_{n-1,n}}V_n
\end{equation}
where all $C_{i,i+1}$ are elliptic curves sharing the same complex structure which we denote by $E$. Moreover $V_0$ and $V_n$ are rational and the other $V_i$'s are ruled over $E$.

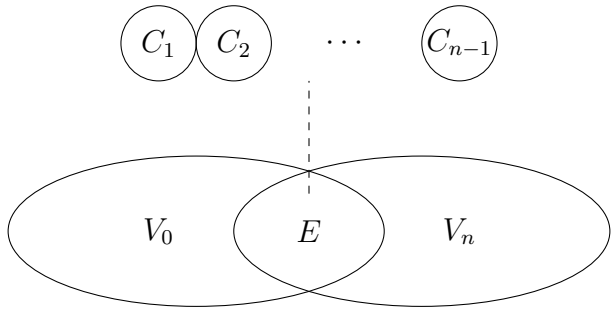
\begin{figure}[h]
\centering
\begin{tikzpicture}
	\draw (-1.5,0) ellipse (2.5cm and 1cm);
	\draw (1.5,0) ellipse (2.5cm and 1cm);
	\node at (0,0) {$E$};
	\node at (-2,0) {$V_0$};
	\node at (2,0) {$V_n$};
	\draw (-2,2.5) ellipse (0.5cm and 0.5cm);
	\draw (-1,2.5) ellipse (0.5cm and 0.5cm);
	\draw (2,2.5) ellipse (0.5cm and 0.5cm);
	\draw[dashed] (0,0.5)--(0,2);
	\node at (-2,2.5) {$C_1$};
	\node at (-1,2.5) {$C_2$};
	\node at (2,2.5) {$C_{n-1}$};
	\node at (0.5,2.5) {$\cdots$};
\end{tikzpicture}\caption{The geometry of $S_p$. $V_i$ is birationally equivalent to $C_i$ fibration over $E$ where $C_i$ is rational. $V_0$ and $V_n$ are rational surfaces.}
\label{fig:geom_manyV}
\end{figure}

The ruled surfaces $V_i$ provide good examples of the conjecture in \cite{2016halverson} as those $V_i$'s admit a fibration structure $\mathbb{P}^1\hookrightarrow V_i\rightarrow E$. After some birational modifications one can assume the elliptic ruled surfaces $V_i$ are minimal and can be contracted along the rulings where $E$ is the sections \cite{1985kondo}. This geometry is now readily recognized as the an $A_{n-1}$ surface singularity over $E$. The geometry of $S_p$ is illustrated in Figure \ref{fig:geom_manyV}.
M2-branes wrapping $C_i$'s in Figure \ref{fig:geom_manyV} will furnish the W-bosons for the enhancement to $SU(n)$. 

Again it is illuminating to take the Kovalev limit. The local physics becomes M-theory on $X\times S^1$ where $X$ is a local CY3 with a compact surface $S_p$. In the limit of collapsing $V_i$'s to $E$ (except for $V_0$ and $V_n$), the W-bosons obtained by M2-brane wrapping $C_i$'s become massless therefore the gauge group is enhanced to $SU(n)$. In the dual picture we expect the 3-3 strings become massless to achieve the same gauge enhancement. Moreover in the M-theory picture we expect there to be an adjoint hyper multiplet since the base of the fibration is a genus one curve \cite{Witten_1996}. Therefore the low energy physics is actually $\mathcal{N}=4$ supersymmetric. In the dual picture we do expect the same amount of SUSY. This is because in the M-theory picture we have only considered the local physics associated with the contractable $A_{n-1}$ surface singularity over $E$ whereas the 7-brane data is encoded in the geometry of $V_0$ and $V_n$, not the $A_{n-1}$ type surface singularity. Therefore in the dual picture the local physics should be described by nothing but $n$ coincident D3-branes filling the flat 4D spacetime without any nearby 7-branes, thus is also $\mathcal{N}=4$.

It is not hard to recognize that this geometry, with $V_n\simeq dP_8$, is actually the same geometry described in \cite{1997ims}. The surface $V_i$, $i=1,\dots,n-1$ are ruled over a genus-1 curve $E$ while $V_n\simeq dP_8$ can be viewed as a $\mathbb{P}_1$-fibration over a genus-0 curve $C$ with 7 reducible fibers \cite{1997douglas}. Moreover $E$ is a double cover of $C$ as discussed in Section~\ref{sec:K3fiber}. Thus in the limit of large base and small fiber the 5D theory obtained from M-theory theory compactification is $Sp(n)+7\mathbf{F}+\mathbf{AS}$. Therefore in the Kovalev limit the 4D $\mathcal{N}=2$ theory is the circle reduction of $Sp(n)+7\mathbf{F}+\mathbf{AS}$.

In the dual picture there is a corresponding $\mathcal{N}=2$ 3/7-brane system. The above contraction of $V_i\rightarrow E$ can be viewed as a mass deformation of the 5D theory hence the 7-brane configuration can again be viewed as decoupling an $I_1$ from the $II^*$ fiber on its $U$-plane as we have mentioned in section~\ref{sec:singlimit} for the single D3-brane probe case. Generalizing the result of \cite{2000hori} we expect the physics to be described by $n$ D3-branes probing an $I^*_3$ singularity. The only difference between this $n$ D3-brane system and the single D3-brane system discussed in section~\ref{sec:singlimit} can be described by the following branching for $n>1$:
\begin{align*}
	U(2n)&\rightarrow Sp(n) \\
	\mathbf{Adj}&\rightarrow \mathbf{Adj}_{n(2n+1)} + \mathbf{AS}_{n(2n-1)}.
\end{align*}
Moreover, together with seven D7-branes there are seven $Sp(n)$ fundamental hypermultiplets. Therefore we see that in the singular limit both the M-theory geometry and D3-brane world volume theory lead to the circle reduction of the 5D $Sp(n)+7\mathbf{F}+\mathbf{AS}$ theory.

It is a well-known fact that the UV completion of 5D $Sp(n)+7\mathbf{F}+\mathbf{AS}$ theory is the 5D rank-$n$ $E_8$ theory \cite{1997extremal}. Therefore it is natural to expect the following\footnote{See Section 3 of \cite{Closset:2020scj} for a detailed discussion of the resolved singular geometry associated with the rank-$n$ $E_8$ theory.}:
\begin{prop}
	The geometry 
	\begin{align*}
		S'_p = V_1\cup_{C_{1,2}}V_2\cup_{C_{2,3}}\cdots\cup_{C_{n-1,n}}V_n
	\end{align*}
	is birationally equivalent to the non-flat fiber obtained from resolving the singular geometry associated with rank-$n$ $E_8$ theory.
\end{prop}

From a 5D point of view, shrinking the surface $S'_p$ corresponds to UV completing 5D $Sp(n)+7\mathbf{F}+\mathbf{AS}$ which leads to the 5D rank-$n$ $E_8$ theory. Further reducing the 5D rank-$n$ $E_8$ theory on a circle leads to 4D rank-$n$ $E_8$ MN theory which can be viewed as $n$ D3-branes probing $II^*$ singularity. The rank-1 and rank-2 cases are well-studied \cite{2021testing, 20215d4d} and we expect this to be true for any $n\in\mathbb{Z}_+$.

Denote by $Z_{E_8^n,-}$ the building block by further tuning $Z_{E_8,sing}$ so that there is $E_8-I_n$ intersection and by $X_{E_8^n,sing}$ the TCS $G_2$ manifold with building blocks $Z_+$ and $Z_{E_8^n,-}$. We formulate the following conjecture generalizing conjecture~\ref{conj:mainconj} for $n$ D3-branes on top of each other:
\begin{conj}
	The following theories are equivalent:
\begin{itemize}[leftmargin=*]
\item
M-theory on $X_{E_8^n,sing}$ in the limit that $n$ surfaces $S_i \subset X_{E_8^n,sing}$ are contracted.
\item
F-theory on $Y_{E_8}$ with $G_4 = 0$ and $n$ $D3$-branes on the $E_8$ singular locus on the base.
\end{itemize}
\end{conj}
Note that dual CY4 geometry is still $Y_{E_8}$ since the CY4 geometry is determined solely by $Z_+$.

\section{Conclusion}

In this work, we have argued that for M-theory compactification on a special class of TCS $G_2$ manifolds $M$, strongly coupled SCFT can be obtained by shrinking a surface $V\subset M$ to a point and it is our main focus to see the $SL(2,\mathbb{Z})$ action on both the F-theory and M-theory side of the duality. Of particular interest is the $SL(2,\mathbb{Z})$ monodromy associated with D3-brane traversing all 7-branes in the system and we find that its M-theory dual action on $X_{dP_n}$ is induced by Serre duality via the tensor product by the canonical bundle on $K_{num}(dP_n)$ in the Kovalev limit. As this construction is local, we also conjecture that its lift in the compact TCS $G_2$ is given by an action on $H^2(X,\mathbb{Z})\oplus H^5(X,\mathbb{Z})$.

Mathematically it will be very interesting to see if the conjectures in this paper can be built upon more rigorous foundations and physically it would be very interesting to see if one can further study the $\mathcal{N}=1$ dynamics directly without going to the Kovalev limit given that the leading order partial SUSY breaking mechanism is a D-term breaking for finite Kovalevton. Though most evidences in this work are from $\mathcal{N}=2$ examples in the Kovalev limit, we expect the partial SUSY breaking at finite Kovalevton do not modify the main conjectures in a drastic way and the study of $\mathcal{N}=1$ dynamics will in turn shed light on the understanding of the geometry of this class of TCS $G_2$ manifolds. 

It will be interesting to further study the deformation of the TCS $G_2$ manifolds $M$, in particular those make $M$ no longer admit a TCS construction, and see how those deformations modify our conjectures on $SL(2,\mathbb{Z})$ monodromies on $X$. As our construction inevitably depends on the fact the 4D theory on the M-theory side is a circle reduction of a 5D $\mathcal{N}=1$ theory, it will be interesting to see how one can decouple the extra $I_1$ that represents the KK modes in the circle reduction. Such a theory with the extra $I_1$ decoupled will in principle be dual to D3-brane probing $E_n$ 7-branes rather than $\widehat{E}_n$ 7-branes therefore is in some sense more interesting for physical applications. 

Moreover, as we have only studied the dual to the monodromy action associated with D3-brane traversing all the 7-branes, it is interesting to study the dual to the monodromy action associated with D3-brane traversing some of the 7-branes in the system. In particular it will be interesting to study the dual of the $SL(2,\mathbb{Z})$ action associated with D3-brane traversing the $E_n$ 7-branes without the extra $I_1$. This will require a more thorough analysis of the general $SL(2,\mathbb{Z})$ monodromies on $G_2$ manifolds and we will study these issues in the future.

\vspace{.5cm}
\noindent \textbf{Acknowledgments.}
We thank Bobby Acharya, Andreas Braun, Mehmet Demirtas, Dave Morrison, and Yi-Nan Wang for helpful discussions. J.T. would like to thank Ying Zhang for her love and support. The work of J.T. is supported by a grant from the Simons Foundation (\#488569, Bobby Acharya). The work of B.S. was partially supported by the NSF Graduate Research Fellowship under grant DGE-1451070 and the ERC Synergy Grant ERC-2020-SyG-854361-HyperK. The work of J.H. is supported
by NSF CAREER grant PHY-1848089.

\appendix

\section{A toric construction of $Z_-$}\label{appendix}

In this appendix we will give a concrete toric model of $Z_{-,sing}$ and its resolution. In particular we will show that there exists a K\"ahler class $J$ that satisfies the four conditions in Section~\ref{sec:contraction} therefore the limit $\text{vol}(V_1) = 0$ can indeed be achieved without destroying the elliptic or $K3$-fibration structure of $Z_-$ which is essential for the M/F-duality to hold.

The toric ambient space $X_{\Delta_{sing}}$ of $Z_{-,sing}$ can be represented by a polytope in the 4D $N$-lattice whose vertices are summarized in the following matrix
\begin{align*}
	\Delta_{sing} = \begin{pmatrix}
		-1 & 0 & 2 & 2 & 2 & 0 & 2 \\
		0 & -1 & 3 & 3 & 3 & 0 & 3 \\
		0 & 0 & -1 & 1 & 0 & 0 & 1 \\
		0 & 0 & 0 & 0 & 1 & -1 & 1
	\end{pmatrix}
\end{align*}
whose columns correspond to the rays $v_x$, $v_y$, $v_{z_1}$, $v_{z_2}$, $v_{\hat{z}_1}$, $v_{\hat{z}_2}$ and $v_{z_e}$ where $v_u$ is the toric ray associated with the toric variable $u$. The singular model $Z_{-,sing}$ can then be constructed as a hyperspace in the toric ambient space and the monomials of its defining equation are given by the polytope in the $M$ lattice whose vertices are 
\begin{align*}
	\nabla = \begin{pmatrix}
		-2 & 1 & 1 & 1 & 1 & 1 & 1 \\
		1 & 1 & 1 & 1 & 1 & -1 & 1 \\
		0 & 1 & 1 & -1 & -1 & 0 & 0 \\
		0 & 0 & -6 & 0 & -5 & 0 & -6
	\end{pmatrix}.
\end{align*} 
It is not hard to check that after suitable coordinate transformation we have a Weierstrass model with
\begin{align}\label{eq:sing_model_2}
	\begin{split}
		&f \propto z_1^4z_2^4, \\
		&g = z_1^5z_2^7z_eP_1^{(6)} + z_1^7z_2^5\hat{z}_1P_2^{(5)} + z_1^6z_2^6Q
	\end{split}
\end{align}
where $\widehat{\mathbb{P}^1}$ is now parameterized by $[\hat{z}_1z_e:\hat{z}_2]$ and $P_i^{(n)}$ is labeled by its degree on $\widehat{\mathbb{P}^1}$. It is easy to see that the $E_8\times E_8$ singular Weierstrass model enhances at $[0:1]$ and another 11 generic points on $\widehat{\mathbb{P}^1}$.

The singular model $Z_{-,sing}$ can be fully resolved by add rays to the polytope $\Delta_{sing}$ whose associated toric variety will be denoted by $X_{\Delta}$. This polytope in the $N$-lattice can be described by the following matrix
\begin{align*}
	\Delta = \begin{pmatrix}
		-1 & 0 & 2 & 2 & 2 & 0 & 2 \\
		0 & -1 & 3 & 3 & 3 & 0 & 3 \\
		0 & 0 & -6 & 6 & 0 & 0 & 1 \\
		0 & 0 & 0 & 0 & 1 & -1 & 1
	\end{pmatrix}
\end{align*}
The resolved $Z_{-,sing}$ is then given by a hypersurface in $X_{\Delta}$ with a fine-regular-star triangulation (FRST) of $\Delta$ whose defining monomials are again given by the polytope $\nabla$ in the $M$-lattice.

In $\Delta$ it is convenient to single out two rays $v_{z_a} = (0,0,1,0)$ and $v_{z_b} = (0,0,-1,0)$. The smooth hypersurface equation of resolved $Z_{-,sing}$ takes the following form
\begin{equation}\label{eq:toric_hypersurface}
	z_az_b\widetilde{P} = z_a^2z_eP_1^{(6)} + z_b^2\hat{z}_1P_2^{(5)}
\end{equation}
where for simplicity we have chosen the same notation for $P_i^{(n)}$ as in Eq.~\ref{eq:sing_model_2} but in general their precise expressions in terms of the toric variables can be different. The SR ideals that will be useful in $X_{\Delta}$, hence in resolved $Z_{-,sing}$, are $z_az_b$, $z_a\hat{z}_1$ and $z_bz_e$ which can easily be checked by giving an arbitrary FRST of $\Delta$.

Let us consider the degeneracy of the $K3$ fiber over point $[0:1]\in\widehat{\mathbb{P}^1}$ in which case both $P_1^{(6)}$ and $P_2^{(5)}$ can be treated as constants and will be denoted by $C_1$ and $C_2$. When $\hat{z}_1 = 0$, we have
\begin{align*}
	z_a\left(z_b\widetilde{P} - C_1z_az_e\right) = 0.
\end{align*}
Due to the SR ideal $z_a\hat{z}_1$, over $[0:1]$ there is one irreducible component when $\hat{z}_1 = 0$ given by
\begin{align*}
	\hat{z}_1 = z_b\widetilde{P} - C_1z_az_e = 0.
\end{align*}
When $z_e = 0$, we have
\begin{align*}
	z_b\left( z_a\widetilde{P} - C_2z_b\hat{z}_1\right) = 0.
\end{align*}
Due to the SR ideal $z_bz_e$, we see again that over $[0:1]\in\widehat{}\mathbb{P}^1$ there is only one irreducible component given by
\begin{align*}
	z_e = z_a\widetilde{P} - C_2z_b\hat{z}_1 = 0.
\end{align*}
Therefore we see that in this slightly modified model over $[0:1]\in\widehat{\mathbb{P}^1}$ the $K3$ fiber splits into two components $\{\hat{z}_1 = 0\}$ and $\{z_e = 0\}$ intersecting the hypersurface in the toric ambient space. 

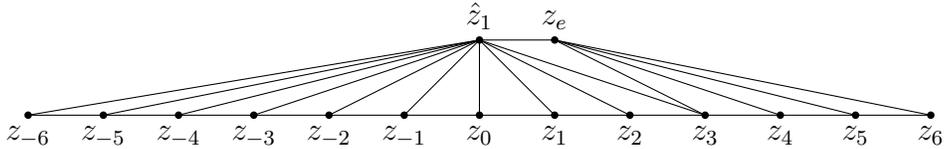
\begin{figure}[h]
	\centering
	\begin{tikzpicture}
		\coordinate[label = above:$\hat{z}_1$] (a) at (0,1);
		\coordinate[label = above:$z_e$] (b) at (1,1);
		\coordinate[label = below:$z_{-6}$] (c6) at (-6,0);
		\coordinate[label = below:$z_{-5}$] (c5) at (-5,0);
		\coordinate[label = below:$z_{-4}$] (c4) at (-4,0);
		\coordinate[label = below:$z_{-3}$] (c3) at (-3,0);
		\coordinate[label = below:$z_{-2}$] (c2) at (-2,0);
		\coordinate[label = below:$z_{-1}$] (c1) at (-1,0);
		\coordinate[label = below:$z_0$] (o) at (0,0);
		\coordinate[label = below:$z_{1}$] (d1) at (1,0);
		\coordinate[label = below:$z_{2}$] (d2) at (2,0);
		\coordinate[label = below:$z_{3}$] (d3) at (3,0);
		\coordinate[label = below:$z_{4}$] (d4) at (4,0);
		\coordinate[label = below:$z_{5}$] (d5) at (5,0);
		\coordinate[label = below:$z_{6}$] (d6) at (6,0);
		\node at (a)[circle,fill,inner sep=1pt]{};
		\node at (b)[circle,fill,inner sep=1pt]{};
		\node at (c6)[circle,fill,inner sep=1pt]{};
		\node at (c5)[circle,fill,inner sep=1pt]{};
		\node at (c4)[circle,fill,inner sep=1pt]{};
		\node at (c3)[circle,fill,inner sep=1pt]{};
		\node at (c2)[circle,fill,inner sep=1pt]{};
		\node at (c1)[circle,fill,inner sep=1pt]{};
		\node at (o)[circle,fill,inner sep=1pt]{};
		\node at (d1)[circle,fill,inner sep=1pt]{};
		\node at (d2)[circle,fill,inner sep=1pt]{};
		\node at (d3)[circle,fill,inner sep=1pt]{};
		\node at (d4)[circle,fill,inner sep=1pt]{};
		\node at (d5)[circle,fill,inner sep=1pt]{};
		\node at (d6)[circle,fill,inner sep=1pt]{};
	\path[every node/.style={auto=false}]
		(a) [-]	edge (b)
		(c6) [-] edge (c5)
		(c5) [-] edge (c4)
		(c4) [-] edge (c3)
		(c3) [-] edge (c2)
		(c2) [-] edge (c1)
		(c1) [-] edge (o)
		(o) [-] edge (d1)
		(d1) [-] edge (d2)
		(d2) [-] edge (d3)
		(d3) [-] edge (d4)
		(d4) [-] edge (d5)
		(d5) [-] edge (d6)
		(a) [-] edge (c6)
		(b) [-] edge (d6)
		(a) [-] edge (c5)
		(a) [-] edge (c4)
		(a) [-] edge (c3)
		(a) [-] edge (c2)
		(a) [-] edge (c1)
		(a) [-] edge (o)
		(a) [-] edge (d1)
		(a) [-] edge (d2)
		(a) [-] edge (d3)
		(b) [-] edge (d3)
		(b) [-] edge (d4)
		(b) [-] edge (d5);
	\end{tikzpicture}
	\caption{The toric fan of the ambient toric variety projected onto the $x_1 = 2$, $x_2 = 3$ plane with a given triangulation. In the diagram $z_0$ represents the ray $v_{z_0} = (2,3,0,0)$.}\label{fig:triangulation}
\end{figure}

Since both $\hat{z}_1$ and $z_e$ are toric variables in this model, it will be relatively easy to look into their properties via the toric diagram. Projecting to the $x_1 = 2$, $x_2 = 3$ (hyper)plane in $\mathbb{C}^4$ we have the triangulation in Figure \ref{fig:triangulation}.

We denote by $S_{\hat{z}_1}$ and $S_{z_e}$ the intersection of the toric divisor $\hat{z}_1 = 0$ and $z_e = 0$ intersecting the hypersurface Eq.~\ref{eq:toric_hypersurface} in the toric ambient space. In this notation the $K3$ fiber $S$ over $[0:1]\in\widetilde{\mathbb{P}^1}$ is $S = S_{\hat{z}_1}\cup S_{z_e}$. For the triangulation in Figure~\ref{fig:triangulation}, $S_{\hat{z}_1}\simeq dP_{12}$ and $S_{z_e}\simeq dP_6$. Clearly by flopping the $(-1)$-curves in $S_{\hat{z}_1}$ and $S_{z_e}$ other $dP_n$'s can also be realized as the shrinking surface and it is manifest that one can do this easily to obtain $n=3,4,5,6,7,8$ by flopping the edges $\hat{z}_1$-$z_{i}$ and $z_e$-$z_i$, e.g., $i=3$ in Figure~\ref{fig:triangulation}.

It is now crucial to check if $S_{z_e}$ can indeed shrink to a point to realize the duality we have conjectured in Section~\ref{sec:singlimit}. To check this we will see that exists a K\"ahler class $J$ such that
\begin{align*}
	J\cdot V_1\cdot V_2\cdot Y = J\cdot V_1\cdot V' = 0
\end{align*}
and
\begin{align*}
	J\cdot V_2\cdot V' \neq 0,\ J\cdot J\cdot V_2 \neq 0, J\cdot J\cdot J\cdot Y\neq 0
\end{align*}
where $V_1 =\{z_4 = 0\}$, $V_2 = \{\hat{z}_1 = 0\}$, $V_{z_3} =\{z_3 = 0\}$ and $Y$ is the class of $Z_-$ in $X_{\Delta}$. It is easy to see that
\begin{align*}
	S_{z_e} = V_1\cdot Y,\ S_{\hat{z}_1} = V_2\cdot Y.
\end{align*}

The toric rays of $\Delta$ are listed in Table~\ref{tab:rays_in_Delta} and the FRST of $\Delta$ we used to obtain a valid $J$ is given in Table~\ref{tab:simplices_dP6}. Concretely we have $V_1 = V_{z_{20}}$, $V_2 = V_{z_{18}}$, $V' = V_{z_{19}}$ and
\begin{align*}
	Y = \sum_{i=1}^{41}V_{z_i} - V_{z_5}
\end{align*}
and the generic $K3$ fiber of $Z_-$ is given by $V_{z_{5}}\cdot Y$. The $(-2)$-curves in $V_1$ associated with the Cartan divisors of $E_6$ are

In this case we choose the basis of the divisors of $X_{\Delta}$ to be
\begin{align*}
	( & 2, 3, 4, 5, 6, 7, 8, 9, 10, 11, 12, 13, 14, 15, 16, 17, 19, 20, 21, 22, \\
	& 23, 24, 26, 27, 29, 30, 31, 32, 33, 34, 35, 36, 37, 38, 39, 40, 41 )
\end{align*}
where the divisors are labeled by their corresponding toric rays in $\Delta$ given by the $\#$ column in Table~\ref{tab:rays_in_Delta}. In terms of this basis we find the following
\begin{align*}
	J = (&4, -12, 42, 0, -4, 14, -4, 23, -8, 28, -11, -10, -8, -6, -4, 0, 4,  \\
	&9, 8, 15, 24, 33, 9, -4, 5, -3, -1, 1, 5, 14, -8, -6, -4, 0, 4, 10, 19)
\end{align*}
in the K\"ahler cone of $X_{\Delta}$ that satisfies all the conditions in Section~\ref{sec:contraction} with the following values
\begin{align*}
	J\cdot V_2\cdot V' = 2,\ J\cdot J\cdot V_2 = 132,\ J\cdot J\cdot J\cdot Y = 2964.
\end{align*}

We have found such $J$'s for $dP_n$ with $n=3,4,5,6,7,8$.

\begin{table}[h]
	\centering
	\begin{tabular}{c|c|c|c|c|c}
	\# & Coordinate & \# & Coordinate & \# & Coordinate \\
	\hline
	1 & $(-1, 0, 0, 0)$ & 14 & $(2, 3, -3, 0)$ & 27 & $(0, 1, -1, 0)$\\
	2 & $(0, -1, 0, 0)$ & 15 & $(2, 3, -2, 0)$ & 28 & $(0, 1, 0, 0)$\\
	3 & $(2, 3, -6, 0)$ & 16 & $(2, 3, -1, 0)$ & 29 & $(0, 1, 1, 0)$\\
	4 & $(2, 3, 6, 0)$ & 17 & $(2, 3, 0, 0)$ & 30 & $(1, 1, -2, 0)$\\
	5 & $(0, 0, 0, -1)$ & 18 & $(2, 3, 0, 1)$ & 31 & $(1, 1, -1, 0)$\\
	6 & $(0, 1, -2, 0)$ & 19 & $(2, 3, 1, 0)$ & 32 & $(1, 1, 0, 0)$\\
	7 & $(0, 1, 2, 0)$ & 20 & $(2, 3, 1, 1)$ & 33 & $(1, 1, 1, 0)$\\
	8 & $(1, 1, -3, 0)$ & 21 & $(2, 3, 2, 0)$ & 34 & $(1, 1, 2, 0)$\\
	9 & $(1, 1, 3, 0)$ & 22 & $(2, 3, 3, 0)$ & 35 & $(1, 2, -3, 0)$\\
	10 & $(1, 2, -4, 0)$ & 23 & $(2, 3, 4, 0)$ & 36 & $(1, 2, -2, 0)$\\
	11 & $(1, 2, 4, 0)$ & 24 & $(2, 3, 5, 0)$ & 37 & $(1, 2, -1, 0)$\\
	12 & $(2, 3, -5, 0)$ & 25 & $(0, 0, -1, 0)$ & 38 & $(1, 2, 0, 0)$\\
	13 & $(2, 3, -4, 0)$ & 26 & $(0, 0, 1, 0)$ & 39 & $(1, 2, 1, 0)$\\
	40 & $(1, 2, 2, 0)$ & 41 & $(1, 2, 3, 0)$
​	\end{tabular}
	\caption{The rays in $\Delta$.}\label{tab:rays_in_Delta}
\end{table}

\begin{longtable}{c|c|c|c|c}
	Simplices & Simplices & Simplices & Simplices & Simplices \\
	\hline
(1, 2, 5, 25) & (2, 5, 30, 31) & (5, 7, 9, 11) & (5, 19, 21, 38) & (13, 14, 18, 36) \\
(1, 2, 5, 26) & (2, 5, 31, 32) & (5, 7, 9, 26) & (5, 21, 22, 33) & (13, 18, 35, 36) \\
(1, 2, 18, 20) & (2, 5, 32, 33) & (5, 7, 11, 41) & (5, 21, 22, 40) & (14, 15, 18, 30) \\
(1, 2, 18, 25) & (2, 8, 18, 25) & (5, 7, 29, 40) & (5, 21, 27, 38) & (14, 15, 18, 37) \\
(1, 2, 20, 26) & (2, 8, 18, 30) & (5, 7, 40, 41) & (5, 21, 27, 39) & (14, 18, 36, 37) \\
(1, 5, 6, 8) & (2, 9, 20, 26) & (5, 8, 12, 13) & (5, 21, 29, 40) & (15, 16, 18, 30) \\
(1, 5, 6, 27) & (2, 9, 20, 34) & (5, 8, 13, 30) & (5, 21, 32, 33) & (15, 16, 18, 37) \\
(1, 5, 7, 26) & (2, 18, 20, 33) & (5, 9, 22, 23) & (5, 22, 40, 41) & (16, 17, 18, 32) \\
(1, 5, 7, 29) & (2, 18, 30, 31) & (5, 9, 22, 34) & (5, 27, 28, 39) & (16, 17, 18, 37) \\
(1, 5, 8, 25) & (2, 18, 31, 32) & (5, 9, 23, 24) & (5, 27, 35, 36) & (16, 18, 30, 31) \\
(1, 5, 21, 29) & (2, 18, 32, 33) & (5, 11, 22, 23) & (5, 27, 36, 37) & (16, 18, 31, 32) \\
(1, 5, 21, 39) & (2, 20, 22, 33) & (5, 11, 22, 41) & (5, 27, 37, 38) & (17, 18, 19, 32) \\
(1, 5, 27, 28) & (2, 20, 22, 34) & (5, 11, 23, 24) & (6, 8, 10, 18) & (17, 18, 19, 38) \\
(1, 5, 28, 39) & (3, 5, 8, 10) & (5, 12, 13, 35) & (6, 10, 18, 35) & (17, 18, 37, 38) \\
(1, 6, 8, 18) & (3, 5, 8, 12) & (5, 13, 14, 30) & (6, 18, 27, 35) & (18, 19, 21, 32) \\
(1, 6, 18, 27) & (3, 5, 10, 35) & (5, 13, 14, 36) & (7, 9, 11, 20) & (18, 19, 21, 38) \\
(1, 7, 20, 26) & (3, 5, 12, 35) & (5, 13, 35, 36) & (7, 9, 20, 26) & (18, 20, 22, 33) \\
(1, 7, 20, 29) & (3, 8, 10, 18) & (5, 14, 15, 30) & (7, 11, 20, 41) & (18, 20, 22, 40) \\
(1, 8, 18, 25) & (3, 8, 12, 18) & (5, 14, 15, 37) & (7, 20, 29, 40) & (18, 20, 29, 40) \\
(1, 18, 20, 29) & (3, 10, 18, 35) & (5, 14, 36, 37) & (7, 20, 40, 41) & (18, 21, 22, 33) \\
(1, 18, 21, 29) & (3, 12, 18, 35) & (5, 15, 16, 30) & (8, 12, 13, 18) & (18, 21, 22, 40) \\
(1, 18, 21, 39) & (4, 5, 9, 11) & (5, 15, 16, 37) & (8, 13, 18, 30) & (18, 21, 27, 38) \\
(1, 18, 27, 28) & (4, 5, 9, 24) & (5, 16, 17, 32) & (9, 20, 22, 23) & (18, 21, 27, 39) \\
(1, 18, 28, 39) & (4, 5, 11, 24) & (5, 16, 17, 37) & (9, 20, 22, 34) & (18, 21, 29, 40) \\
(2, 5, 8, 25) & (4, 9, 11, 20) & (5, 16, 30, 31) & (9, 20, 23, 24) & (18, 21, 32, 33) \\
(2, 5, 8, 30) & (4, 9, 20, 24) & (5, 16, 31, 32) & (11, 20, 22, 23) & (18, 27, 28, 39) \\
(2, 5, 9, 26) & (4, 11, 20, 24) & (5, 17, 19, 32) & (11, 20, 22, 41) & (18, 27, 35, 36) \\
(2, 5, 9, 34) & (5, 6, 8, 10) & (5, 17, 19, 38) & (11, 20, 23, 24) & (18, 27, 36, 37) \\
(2, 5, 22, 33) & (5, 6, 10, 35) & (5, 17, 37, 38) & (12, 13, 18, 35) & (18, 27, 37, 38) \\
(2, 5, 22, 34) & (5, 6, 27, 35) & (5, 19, 21, 32) & (13, 14, 18, 30) & (20, 22, 40, 41)\\
\caption{Simplices of the chosen FRST of $\Delta$. Each 4D cone of the FRST of $\Delta$ is expanded by the rays $(v_i,v_j,v_k,v_l)$ for the simplex $(i,j,k,l)$.}
\label{tab:simplices_dP6}
\end{longtable}

\newpage
\bibliographystyle{JHEP}
\bibliography{G2}

\end{document}